# Physics-augmented models to simulate commercial adaptive cruise control (ACC) systems

Yinglong He [a], Marcello Montanino [b,] *, Konstantinos Mattas [c], Vincenzo Punzo [b], Biagio Ciuffo [c]

[a] University of Birmingham, Edgbaston, Birmingham, B15 2TT, UK
[b] University of Naples Federico II, Via Claudio, 21, 80125 Napoli, IT
[c] European Commission – Joint Research Centre., Ispra (VA), IT

* Corresponding Author: Ph.: +39 081 7683770

E-mail addresses:

yxh701@student.bham.ac.uk
marcello.montanino@unina.it
konstantinos.mattas@ec.europa.eu
vincenzo.punzo@unina.it
biagio.ciuffo@ec.europa.eu

# Abstract

This paper investigates the accuracy and robustness of car-following (CF) and adaptive cruise control (ACC) models used to simulate measured driving behaviour of commercial ACCs. To this aim, a general modelling framework is proposed, in which ACC and CF models have been incrementally augmented with physics extensions; namely, perception delay, linear or nonlinear vehicle dynamics, and acceleration constraints. The framework has been applied to the Intelligent Driver Model (IDM), the Gipps' model, and to three basic ACCs. These are a linear controller coupled with a constant time-headway spacing policy and with two other policies derived from the traffic flow theory, which are the IDM desired-distance function and the Gipps' equilibrium distance-speed function. The ninety models resulting from the combination of the five base models and the aforementioned physics extensions, have been assessed and compared through a vast calibration and validation experiment against measured trajectory data of low-level automated vehicles. When a single extension has been applied, perception delay and linear dynamics have been the extensions to mostly increase modelling accuracy, whatsoever the base model considered. Concerning models, Gipps'-based ones have outperformed all other CF and ACC models in calibration. Even among ACCs, the linear controllers coupled with a Gipps' spacing policy have been the best performing. On the other hand, IDM-based models have been by far the most robust in validation, showing almost no crash when calibrated parameters have been used to simulate different trajectories. Overall, the paper shows the importance of cross-fertilization between traffic flow and vehicle studies.

**Highlights:**

1. We compared the accuracy of ninety models, among CF models and ACC algorithms.
2. New controllers with spacing policies from the IDM and Gipps' models are studied.
3. The impact on accuracy of increasingly complex lower-level dynamics is evaluated.
4. ACC trajectories from platoon experiments with five high-end vehicles are applied.
5. Gipps'-based models are the most accurate in reproducing measured ACC dynamics.

# 1. Introduction

Accurate simulation of mixed traffic, including human-driven vehicles (HDVs) and automated vehicles (AVs), is crucial to the development of cooperative, connected, and automated mobility (CCAM) technologies and to the assessment of their impacts on traffic efficiency and the related externalities (Alonso Raposo et al., 2019). Car-following models (CF) and adaptive cruise control algorithms (ACC) have been proposed to mimic dynamics of HDVs and AVs, respectively.

A vast literature exists on modelling HDVs through behavioural car-following models. Since the true underlying human driving process is unknown, several formulations coexist in the field literature, though based on completely different assumptions. If theoretical properties, advantages, and drawbacks of these models have been confronted and are very well known, it has been never demonstrated the dominance of one model over the others in terms of performances, i.e., as to the distance between simulated and observed HDV trajectories. Previous studies have shown that, when calibrating models against trajectories, modelling errors vary from one trajectory to another more than from one model to another. That is to say, there are trajectories which are reasonably reproduced by most of the models, and others which all models fail to mimic at an acceptable degree of accuracy (Punzo and Simonelli, 2005; Punzo et al., 2021). Basic CF models have been applied also to emulate the dynamics of ACC vehicles, though these models were rarely calibrated against field trajectories (Darbha et al., 1994; Milanés and Shladover, 2014; Gunter et al., 2019; Wang et al., 2021). In general, behavioural CF models barely consider detailed lower-level dynamics concerning the characteristics of the vehicle, the sensors, the road, etc. In fact, they are conceived as stimulus-response relationships, which phenomenologically describe the vehicle motion, without making the vehicle control loop and the vehicle dynamics explicit (Rothery, 2001).

On the other hand, a plethora of ACC algorithms have been studied and proposed for implementation in AVs (for a review, see Yu et al., 2021). Most of them are coupled with lower-level dynamics at different levels of detail (e.g., Li et al., 2017; He et al., 2019). Opposite to studies on behavioural CF models, calibration of ACC algorithms against observed trajectories has not generally been a subject of investigation, since main objective of ACC studies has been the design of automated logics to replace the human one, and not to mimic it (an example of ACC calibration can be found in Wang et al., 2021). ACC algorithms proposed in the literature have been also widely applied to represent AVs behaviour in studies attempting to assess the impact of AVs on the efficiency and safety of mixed traffic (e.g., Shladover et al., 2012; Talebpour and Mahmassani, 2016; Mattas et al., 2018). Being ACC algorithms not usually calibrated on observed AVs trajectories, the extent to which the dynamics they describe are close to those produced by the undisclosed ACC algorithms of commercial vehicles, is unknown.

If we neglect lower-level dynamics, behavioural CF models and ACC algorithms are conceptually equivalent. In fact, behavioural CF models can be treated as upper-level controllers to all effects (like ACC algorithms), as they provide the ego vehicle with the desired acceleration or speed, as a function of its own and preceding vehicle kinematics. Conversely, we may think of ACC models as elementary CF models.

Therefore, from a modelling standpoint, there is no true reason to prefer CF models to simulate HDVs and ACC algorithms to simulate AVs. In fact, when simulating a mixed traffic, we retain almost the same level of ignorance about the actual dynamics of HDVs and AVs.

On the one hand, we do not know the true logic that inspires human driver behaviours as well as the true controller of AVs (unless all manufacturers disclose their proprietary algorithms). On the other hand, we acknowledge a heterogeneity of behaviours among human drivers (see Punzo and Montanino, 2020), as well as a heterogeneity of controllers among AV manufacturers, and of algorithm settings among AV drivers (each driver may select a preferred driving “style”, that is more or less aggressive time-headway settings).

In a modelling view, the only true difference between HDVs and AVs, is that the logic of AVs is not expected to be stochastic in time, opposite to HDVs, see e.g., the stochastic variability of human drivers’ reaction times (this consideration has motivated a number of studies investigating stochastic behavioural CF models, see e.g., Treiber and Kesting, 2018; Tian et al., 2019).

Therefore, once acknowledged that behavioural CF models and ACC algorithms are conceptually equivalent, a number of questions arise.

First, one may wonder what the impact is on behavioural CF model performances, of (considering CF models as upper-level controllers and) augmenting them with detailed lower-level dynamics, which account for perception delays, actuation lags, vehicle and motor constraints (see e.g. Treiber et al., 2006; Makridis et al., 2019; He et al., 2020a).

Second, one may wonder what the ability of ACC algorithms from the field literature to capture actual dynamics of commercial ACC vehicles, is. Since the literature on ACCs is puzzled with lower-level dynamics at different levels of detail, a systematic investigation is necessary to assess the contribution of each physical extension to the accuracy of simulation results, for ACC algorithms too.

Eventually, one might compare performances of behavioural CF models and ACC algorithms, once augmented with the same physical extensions, and verify whether significant differences emerge from the two classes of models (recently, CF models were calibrated on ACC trajectories to simulate ACC behaviour, though no physical extensions were considered there; see Wang et al., 2021; De Souza and Stern, 2021).

To answer these questions, this paper develops a comparison framework that allows confronting different models, each one augmented with physical extensions of increasing complexity. To this end, all possible combinations of five base models (CF and ACC) with different extensions, including perception delay, linear or nonlinear vehicle dynamics, and acceleration constraints, are calibrated against measured vehicle trajectory data.

As base models, two CF models, namely the IDM (Treiber et al., 2000) and Gipps' model (Gipps, 1981), and three ACC algorithms have been tested. Applied ACCs consist of a linear controller (e.g., Li et al., 2017) coupled with a Constant Time Headway (CTH) spacing policy (e.g., Zheng et al., 2016), and with two other policies derived from the traffic flow theory; namely, the IDM desired distance, and the Gipps' equilibrium distance-speed function. To the best of our knowledge, the approach of coupling a linear control law with spacing policies derived from the car-following theory is novel in the literature.

As a result, 90 different models (i.e., 5 base models per 18 variants each) have been calibrated, each one against open-source trajectories gathered on 7 platooning experiments with four high-end vehicles controlled by ACC systems (Audi, Tesla, Mercedes, and BMW), following a platoon leader (Makridis et al., 2021). It has been chosen to adopt trajectories from AVs – and not from HDVs – since HDVs trajectories reflect time stochasticity of human drivers' behaviours, which would have puzzled analysis results. Models calibrated against a vehicle trajectory in a platoon, have been also cross validated against trajectories of the same vehicle taken from other platoons.

By calibrating models against trajectories from automated driving, it has been also possible to verify whether the consolidated habit of applying linear control laws – and not CF models – to simulate ACC vehicles has significance in practice.

The remainder of the paper is organized as follows. Section 2 describes the unified modelling framework adopted in this study, which includes five base models and three types of physics-extensions. Section 3 describes the field platoon data applied for model calibration and validation. Section 4 provides the methodology and the design of experiments. Sections 5 and 6 discuss the calibration and validation results, respectively. Afterwards, Section 7 draws the conclusion.

# 2. Modelling framework

The modelling framework applied in this work is presented in Fig. 1. The framework is applied to both behavioural CF models and ACC algorithms. Through the framework, we also clarified the equivalence, from a modelling standpoint, between behavioural CF models and ACC algorithms, and we showed how both types of models can be augmented with lower-level dynamics of increasing complexity.

The five base models studied here are the IDM and Gipps' CF model, and three ACC algorithms that adopt a linear control law coupled either with a CTH spacing policy (L-CTH), and with two new spacing policies derived from the IDM and Gipps' models, L-IDM and L-Gipps respectively. Each base model was formulated in 18 variants, which account for all possible combinations of a base model formulation with physics-enhancements, including perception delay, vehicle dynamics and acceleration constraints, at increasing level complexity.

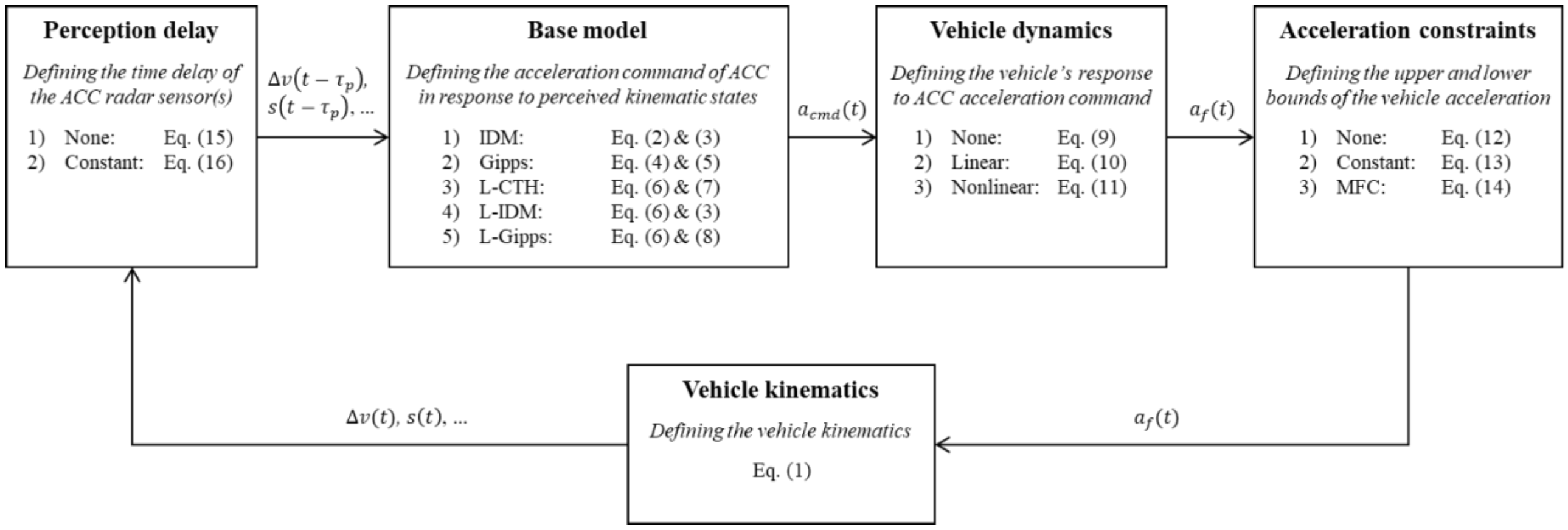

**Fig. 1. Unified modelling framework.**

The base model returns the acceleration command $a_{cmd}(t)$ to be executed by the ego vehicle *f* at time *t* as a function of model parameters and the following inputs: the ego vehicle speed $v_f(t-\tau_p)$, the inter-vehicle spacing $s(t-\tau_p) = x_l(t-\tau_p) - x_f(t-\tau_p) - L$ and the relative speed $\Delta v(t-\tau_p) = v_l(t-\tau_p) - v_f(t-\tau_p)$ from a preceding vehicle *l*, where $x_i$ denotes the longitudinal position of vehicle *i*, *L* the preceding vehicle length and $\tau_p$ a perception delay (for $\tau_p$, see Section 2.2.3). The acceleration command is then transformed into a control signal acting on the dynamic system of the ego vehicle (Section 2.2.1). Acceleration constraints are eventually applied, providing the vehicle acceleration ($a_f(t)$) (these constraints are due, for instance, to limited motor acceleration capabilities; see Section 2.2.2).

Ego vehicle kinematics are thus obtained by coupling the CF/ACC model with the equation of motion. In numerical integration, a ballistic integration scheme is typically adopted (Treiber et al., 2006; Treiber and Kanagaraj, 2015):

$$\begin{cases} v_f(t) = v_f(t-\Delta t) + a_f(t)\cdot\Delta t \\ x_f(t) = x_f(t-\Delta t) + \dfrac{v_f(t) + v_f(t-\Delta t)}{2}\cdot\Delta t \end{cases} \tag{1}$$

where $\Delta t$ is the numerical integration step.

## 2.1. Base models

In the proposed modelling framework, the base model defines the acceleration command ($a_{cmd}$) to the ACC vehicle, in response to perceived kinematic states, i.e., $s$ and $\Delta v$. The five base models studied in this work are presented in the next sub-sections.

### 2.1.1. IDM model

The Intelligent Driver Model (IDM) is developed by Treiber et al. (2000) and is widely investigated in the traffic flow theory literature (see e.g., Ward and Wilson, 2011; Treiber and Kesting, 2011; Montanino et al., 2021; Montanino and Punzo, 2021):

$$a_{cmd}(t) = a_{max}\cdot\left(1 - \left(\frac{v_f(t)}{v_0}\right)^{\delta} - \left(\frac{s_{des}(t)}{s(t)}\right)^2\right) \tag{2}$$

where $a_{cmd}$ is the acceleration command to be applied by the ego vehicle, i.e., transformed into control signals of the throttle and brake pedals; $a_{max}$ indicates the maximum acceleration rate; $v_0$ is the desired speed, which represents the set speed of the onboard ACC system in this work; $s_{des}$ indicates the desired inter-vehicle spacing; $\delta$ is an exponent factor.

The desired spacing policy ($s_{des}$) in the IDM model is calculated as:

$$s_{des}(t) = d_0 + \max\left(0, t_h v_f(t) - \frac{v_f(t)\Delta v(t)}{2\sqrt{-a_{max}a_{min}}}\right) \tag{3}$$

where $d_0$ is the minimum desired (or standstill) spacing; $t_h$ denotes the desired time headway; $a_{min}$ is the comfort deceleration rate (negative).

### 2.1.2. Gipps' model

Gipps' model (Gipps, 1981) specifies a safe inter-vehicle spacing and plans the vehicle speed for following instants (for a review see Ciuffo et al., 2012). Therefore, the acceleration command ($a_{cmd}$) is derived from the speed planned for the following instant and the speed at the current time, as defined by:

$$a_{cmd}(t) = \left(v_{cmd}(t + t_h) - v_f(t)\right)/t_h \tag{4}$$

where $t_h$ is an "apparent" reaction time, and $v_{cmd}$ denotes the safe speed planned by the Gipps' model that considers two driving regimes. The former is a free-flow regime where a vehicle accelerates to its desired speed, following a pre-defined acceleration pattern (a generalized acceleration pattern can be found in Ciuffo et al., 2012). The latter is a car-following regime where the vehicle plans a speed to prevent rear-end collision even in the occurrence of a sudden deceleration of the preceding vehicle. The vehicle chooses the minimum between the speed values resulting from the two regimes:

$$v_{cmd}(t + t_h) = \min\left(\begin{array}{c} v_f(t) + 2.5a_{max}t_h \cdot \left(1 - \frac{v_f(t)}{v_0}\right)\left(0.025 + \frac{v_f(t)}{v_0}\right)^{0.5}, \\ a_{min} \cdot \left(\frac{t_h}{2} + \theta\right) + \sqrt{a_{min}{}^2 \cdot \left(\frac{t_h}{2} + \theta\right)^2 - a_{min} \cdot \left(2(s(t) - d_0) - t_h v_f(t) - \frac{v_l(t)^2}{\hat{a}_{min}}\right)} \end{array}\right) \tag{5}$$

where $a_{max}$ indicates the maximum acceleration rate; $a_{min}$ is the comfort deceleration rate (negative); $\hat{a}_{min}$ is the following vehicle estimate of preceding vehicle's deceleration rate (negative); $d_0$ is the minimum desired spacing (at standstill); $v_0$ is the desired speed; $\theta$ is an additional delay inserted to avoid braking always at the maximum rate.

### 2.1.3. Linear control law

Linear control laws are commonly used in studies of ACC and cooperative ACC, as these models can significantly simplify theoretical analyses (Zheng et al., 2016; Li et al., 2017; He et al., 2020c). In this work, we studied three control algorithms, which couple the linear control law with different spacing policies. The general formulation of the linear control law is the following:

$$\begin{cases} a_{cmd}(t) = \min\left\{k_v \cdot \Delta v(t) - k_s \cdot \Delta s(t), k_0 \cdot \left(v_0 - v_f(t)\right)\right\} \\ \Delta s(t) = s_{des}(t) - s(t) \end{cases} \tag{6}$$

where $s_{des}$ is the desired inter-vehicle spacing; $\Delta s$ denotes the error between the desired ($s_{des}$) and the actual ($s$) spacing values; $k_s$ is the gain factor for the spacing error; $k_v$ is the gain factor for the speed difference between the preceding vehicle and the ego vehicle; $k_0$ is the gain factor for the error between the ACC set speed ($v_0$) and ego vehicle speed ($v_f$).

The desired spacing is determined by a spacing policy. Many researchers have studied spacing policies from microscopic or macroscopic perspectives, aiming to improve e.g., car-following safety, driver acceptance, stability performance (see e.g., Zhang et al., 2019; Wu et al., 2020). For comparison purposes, in this study, three spacing policies are coupled with the linear control law in Eq. (6).

First, a constant time headway (CTH) spacing policy calculates the desired spacing as a linear function of the ego vehicle speed ($v_f$), in which the slope and the intercept of the linear function are the time headway ($t_h$) and the standstill spacing ($d_0$), respectively, i.e.,

$$s_{des}(t) = d_0 + t_h v_f(t) \tag{7}$$

In addition, two new spacing policies are tested in this work. They are referred to as IDM-desired and Gipps-equilibrium spacing policies, since they are derived from the IDM and Gipps' model, respectively. These two spacing policies for a linear controller are a novel contribution to the ACC field literature. The former is given by Eq. (3) as per the IDM. The latter implements the distance-speed function at equilibrium of Gipps' model, which has been derived in Punzo and Tripodi (2007):

$$s_{des}(t) = d_0 + (t_h + \theta)v_f(t) - 0.5v_f(t)^2\left(\frac{1}{a_{\min}} - \frac{1}{\hat{a}_{\min}}\right) \tag{8}$$

For simplicity, the linear control law, i.e., Eq. (6), coupled with the spacing policies described in Eq. (7), (3), and (8), are hereafter abbreviated as L-CTH, L-IDM, and L-Gipps, respectively.

## 2.2. Physics extensions

Most of the studies on behavioural CF models do not consider lower-level dynamics (with some exceptions, e.g., He et al., 2020a). On the contrary, most of ACC studies account for these dynamics at different levels of detail. The resulting literature is puzzled, with three major extensions, namely, vehicle dynamics (e.g., actuation lag), acceleration constraints, and perception delay, variously applied.

Two types of vehicle dynamics, i.e., linear and nonlinear, are applied in the literature, both striking a balance between simplicity and accuracy (Xiao et al., 2017). Linear vehicle dynamics introduces the first-order lag between the acceleration command generated by the ACC and the actual acceleration of the ego vehicle, in order to represent the response of the driveline system (VanderWerf et al., 2001; Jia and Ngoduy, 2016), whereas the effects of external factors cannot be captured. On the other hand, nonlinear vehicle dynamics can simulate both internal (e.g., driveline) and external (e.g., resistances of aerodynamic, rolling, and grade) factors (e.g., Zheng et al., 2017), in which a variable road slope can therefore be considered.

Acceleration constraints are specified to avoid unrealistic accelerations beyond the vehicle's practical capabilities. In most models, the upper and lower acceleration bounds are either set to constant values or not implemented. For example, the IDM only has a constant upper bound for acceleration ($a_{max}$), and no lower bound value, thus potentially leading to unrealistic large deceleration when the inter-vehicle spacing drops significantly (e.g., in cut-in manoeuvres; Kesting et al., 2010). Similar issues are observed with Gipps' model.

In the real world, upper and lower acceleration bounds are not constant and are affected by many factors such as the engine, the braking system, and the road loads (Mannering and Washburn, 2020). Consequently, many newly proposed CF models try to account for adaptive acceleration constraints that vary as functions of speed, e.g., Rakha (2009), Fadhloun and Rakha (2020), and the microsimulation free-flow acceleration (MFC) model (Makridis et al., 2019). For a review see He et al. (2020a).

Perception delay represents the dynamics of human senses or automated sensors in data gathering and processing. It plays an important role in traffic properties, e.g., string stability (Xiao and Gao, 2011). For example, experimental studies exhibit that linear ACC laws with perception delay can accurately capture the string instability of ACC platoons in field tests (e.g., Gunter et al., 2020). Concerning CF models, the IDM model, for instance, has been enhanced to include the perception delay (e.g., Treiber et al., 2006).

In the following subsections, the formulations of the aforementioned physical extension types, which are studied in this paper, are presented.

**Remark.** When behavioural CF models are considered as upper-level controllers, and are augmented with lower-level dynamics, their descriptive capability might decrease, i.e., the CF model ability of reproducing observed vehicle trajectories decreases. In fact, adding e.g., nonlinear dynamics or motor constraints, makes the simulated vehicle motion different from that returned by the original CF model, i.e., provides an output which is inconsistent with the original model assumptions and calculations.

### 2.2.1. Vehicle dynamics

Longitudinal vehicle dynamics define how the vehicle responds to the acceleration command generated by the ACC (Li et al., 2017). The types of vehicle dynamics applied in this study can be categorized as follows.

1) None: No vehicle dynamics is applied, i.e., the resulting ego vehicle acceleration coincides with the ACC acceleration command. Therefore, it is the simplest case, and is described by:

$$a_f(t) = a_{cmd}(t) \tag{9}$$

2) Linear: It captures the actuation lag ($\tau_a$) of the driveline system, and is expressed as:

$$\tau_a \dot{a_f}(t) + a_f(t) = a_{cmd}(t) \tag{10}$$

3) Nonlinear: It reproduces nonlinear features, including internal driveline system and external road loads. The corresponding vehicle dynamics are depicted by the following equations:

$$
\begin{cases}
\tau_a \, \dot{a}_t(\mathrm{t}) + a_t(\mathrm{t}) = a_{cmd}(\mathrm{t}) \\
F_t(t) = \dfrac{T_t(t)}{r_w} = m_f a_t(\mathrm{t}) \\
F_r(t) = f_0 \cos\alpha(t) + f_1 v_e(t) + f_2 v_e(t)^2 + m_f g \sin\alpha(\mathrm{t}) \\
a_f(t) = \dfrac{F_t(t) - F_r(t)}{\varphi_f m_f}
\end{cases}
\tag{11}
$$

where $F_r$ is the total resistance force, which includes aerodynamics, rolling, and grade resistances; $f_0$, $f_1$, and $f_2$ are road load coefficients (He et al., 2020b); $\alpha$ is the road grade; $g$ is the gravitational acceleration; $F_t$, $T_t$, and $a_t$ are the tractive force, the tractive torque, and the acceleration, respectively, acting on the wheels; $r_w$ is the wheel radius; $\tau_a$ is the actuation lag; $\varphi_e$ is the equivalent inertial mass factor; $m_f$ represent the operating mass, and is equal to the sum of the unladen mass ($m_0$) and the mass of vehicle loads ($m_{load}$) (e.g., passengers and measuring instruments).

### 2.2.2. Acceleration constraints

Acceleration constraints define the maximum capabilities of the vehicle propulsion and braking systems in various conditions. Three types of acceleration constraints are studied:

1) None: no physical constraint is applied to the ego vehicle acceleration, i.e.:

$$-\infty \leq a_f(t) \leq +\infty \tag{12}$$

2) Constant: it assumes that the ego vehicle acceleration is always bounded by constant values, as described by:

$$a_{lb} \leq a_f(t) \leq a_{ub} \tag{13}$$

where $a_{lb}$ and $a_{ub}$ are respectively the lower and upper bounds of the ego vehicle acceleration (set to -7 and 5 $m/s^2$ in this work).

3) Microsimulation free-flow acceleration (MFC) boundary model: the MFC model serves to define dynamic bounds for the ego vehicle acceleration, in which both lower and the upper bounds change with the vehicle speed $v_f(t)$. The constraints are described by:

$$a_{dp}\left(v_f(t)\right) \leq a_f(t) \leq a_{ap}\left(v_f(t)\right) \tag{14}$$

where $a_{ap}$ and $a_{dp}$ are nonlinear curves representing acceleration and deceleration potentials of the ego vehicle, respectively. These curves are based on previous studies by Makridis et al. (2019) and He et al. (2020a).

### 2.2.3. Perception delay

Perception delay is a significant phenomenon affecting AV dynamics, and is caused by sensors and/or V2X technologies. To capture characteristics of ACC radar sensors, the parameter $\tau_p$ is introduced in the modelling framework. It represents the time delay of sensor perception. Consequently, the time instance $t$ in all state variables and inputs in Eq. (2), (3), and (5)-(8), i.e., in $v_f(t)$, $v_l(t)$, $\Delta v(t)$, $s(t)$, $s_{des}(t)$, and $\Delta s(t)$, is replaced with the delayed time instance $t - \tau_p$. This study provides two options of perception delay as follows.

1) None:

$$\tau_p = 0 \tag{15}$$

2) Constant:

$$\tau_p > 0 \tag{16}$$

### 2.2.4. Model summary

The full list of tested models is summarized in Table 1. Ninety models are derived by combining the five base models with three vehicle dynamics formulations, three acceleration constraint functions, and two perception delay types, in a full factorial design. In the table, model variants relative to the same base model are sorted by increasing level of complexity.

# Table 1. Configurations of investigated models.

| Model groups | Model ID | ACC controller | Spacing policy | Perception delay (PD) | Vehicle dynamics (VD) | Acceleration Constraints (AC) | Calibration parameters |
|---|---|---|---|---|---|---|---|
| IDM-based models | 1 [a] | IDM | IDM-desired | None | None | None | $\delta, v_0, d_0, t_h, a_{max}, a_{min}$ |
| | 2 | IDM | IDM-desired | None | None | Constant | $\delta, v_0, d_0, t_h, a_{max}, a_{min}$ |
| | 3 | IDM | IDM-desired | None | None | MFC | $\delta, v_0, d_0, t_h, a_{max}, a_{min}$ |
| | 4 | IDM | IDM-desired | None | Linear | None | $\delta, v_0, d_0, t_h, a_{max}, a_{min}, \tau_a$ |
| | 5 | IDM | IDM-desired | None | Linear | Constant | $\delta, v_0, d_0, t_h, a_{max}, a_{min}, \tau_a$ |
| | 6 | IDM | IDM-desired | None | Linear | MFC | $\delta, v_0, d_0, t_h, a_{max}, a_{min}, \tau_a$ |
| | 7 | IDM | IDM-desired | None | Nonlinear | None | $\delta, v_0, d_0, t_h, a_{max}, a_{min}, \tau_a, m_{load}, f_0, f_1, f_2$ |
| | 8 | IDM | IDM-desired | None | Nonlinear | Constant | $\delta, v_0, d_0, t_h, a_{max}, a_{min}, \tau_a, m_{load}, f_0, f_1, f_2$ |
| | 9 | IDM | IDM-desired | None | Nonlinear | MFC | $\delta, v_0, d_0, t_h, a_{max}, a_{min}, \tau_a, m_{load}, f_0, f_1, f_2$ |
| | 10 | IDM | IDM-desired | Constant | None | None | $\delta, v_0, d_0, t_h, a_{max}, a_{min}, \tau_p$ |
| | 11 | IDM | IDM-desired | Constant | None | Constant | $\delta, v_0, d_0, t_h, a_{max}, a_{min}, \tau_p$ |
| | 12 | IDM | IDM-desired | Constant | None | MFC | $\delta, v_0, d_0, t_h, a_{max}, a_{min}, \tau_p$ |
| | 13 | IDM | IDM-desired | Constant | Linear | None | $\delta, v_0, d_0, t_h, a_{max}, a_{min}, \tau_p, \tau_a$ |
| | 14 | IDM | IDM-desired | Constant | Linear | Constant | $\delta, v_0, d_0, t_h, a_{max}, a_{min}, \tau_p, \tau_a$ |
| | 15 | IDM | IDM-desired | Constant | Linear | MFC | $\delta, v_0, d_0, t_h, a_{max}, a_{min}, \tau_p, \tau_a$ |
| | 16 | IDM | IDM-desired | Constant | Nonlinear | None | $\delta, v_0, d_0, t_h, a_{max}, a_{min}, \tau_p, \tau_a, m_{load}, f_0, f_1, f_2$ |
| | 17 | IDM | IDM-desired | Constant | Nonlinear | Constant | $\delta, v_0, d_0, t_h, a_{max}, a_{min}, \tau_p, \tau_a, m_{load}, f_0, f_1, f_2$ |
| | 18 | IDM | IDM-desired | Constant | Nonlinear | MFC | $\delta, v_0, d_0, t_h, a_{max}, a_{min}, \tau_p, \tau_a, m_{load}, f_0, f_1, f_2$ |
| Gipps-based models | 19 [a] | Gipps | *inherent* | None | None | None | $\theta, v_0, d_0, t_h, a_{max}, a_{min}, \hat{a}_{min}$ |
| | 20 | Gipps | *inherent* | None | None | Constant | $\theta, v_0, d_0, t_h, a_{max}, a_{min}, \hat{a}_{min}$ |
| | 21 | Gipps | *inherent* | None | None | MFC | $\theta, v_0, d_0, t_h, a_{max}, a_{min}, \hat{a}_{min}$ |
| | 22 | Gipps | *inherent* | None | Linear | None | $\theta, v_0, d_0, t_h, a_{max}, a_{min}, \hat{a}_{min}, \tau_a$ |
| | 23 | Gipps | *inherent* | None | Linear | Constant | $\theta, v_0, d_0, t_h, a_{max}, a_{min}, \hat{a}_{min}, \tau_a$ |
| | 24 | Gipps | *inherent* | None | Linear | MFC | $\theta, v_0, d_0, t_h, a_{max}, a_{min}, \hat{a}_{min}, \tau_a$ |
| | 25 | Gipps | *inherent* | None | Nonlinear | None | $\theta, v_0, d_0, t_h, a_{max}, a_{min}, \hat{a}_{min}, \tau_a, m_{load}, f_0, f_1, f_2$ |
| | 26 | Gipps | *inherent* | None | Nonlinear | Constant | $\theta, v_0, d_0, t_h, a_{max}, a_{min}, \hat{a}_{min}, \tau_a, m_{load}, f_0, f_1, f_2$ |
| | 27 | Gipps | *inherent* | None | Nonlinear | MFC | $\theta, v_0, d_0, t_h, a_{max}, a_{min}, \hat{a}_{min}, \tau_a, m_{load}, f_0, f_1, f_2$ |
| | 28 | Gipps | *inherent* | Constant | None | None | $\theta, v_0, d_0, t_h, a_{max}, a_{min}, \hat{a}_{min}, \tau_p$ |
| | 29 | Gipps | *inherent* | Constant | None | Constant | $\theta, v_0, d_0, t_h, a_{max}, a_{min}, \hat{a}_{min}, \tau_p$ |
| | 30 | Gipps | *inherent* | Constant | None | MFC | $\theta, v_0, d_0, t_h, a_{max}, a_{min}, \hat{a}_{min}, \tau_p$ |
| | 31 | Gipps | *inherent* | Constant | Linear | None | $\theta, v_0, d_0, t_h, a_{max}, a_{min}, \hat{a}_{min}, \tau_p, \tau_a$ |
| | 32 | Gipps | *inherent* | Constant | Linear | Constant | $\theta, v_0, d_0, t_h, a_{max}, a_{min}, \hat{a}_{min}, \tau_p, \tau_a$ |
| | 33 | Gipps | *inherent* | Constant | Linear | MFC | $\theta, v_0, d_0, t_h, a_{max}, a_{min}, \hat{a}_{min}, \tau_p, \tau_a$ |
| | 34 | Gipps | *inherent* | Constant | Nonlinear | None | $\theta, v_0, d_0, t_h, a_{max}, a_{min}, \hat{a}_{min}, \tau_p, \tau_a, m_{load}, f_0, f_1, f_2$ |
| | 35 | Gipps | *inherent* | Constant | Nonlinear | Constant | $\theta, v_0, d_0, t_h, a_{max}, a_{min}, \hat{a}_{min}, \tau_p, \tau_a, m_{load}, f_0, f_1, f_2$ |
| | 36 | Gipps | *inherent* | Constant | Nonlinear | MFC | $\theta, v_0, d_0, t_h, a_{max}, a_{min}, \hat{a}_{min}, \tau_p, \tau_a, m_{load}, f_0, f_1, f_2$ |
| L-CTH-based models | 37 [a] | Linear contr. | CTH | None | None | None | $k_s, k_v, k_0, v_0, d_0, t_h$ |
| | 38 | Linear contr. | CTH | None | None | Constant | $k_s, k_v, k_0, v_0, d_0, t_h$ |
| | 39 | Linear contr. | CTH | None | None | MFC | $k_s, k_v, k_0, v_0, d_0, t_h$ |
| | 40 | Linear contr. | CTH | None | Linear | None | $k_s, k_v, k_0, v_0, d_0, t_h, \tau_a$ |
| | 41 | Linear contr. | CTH | None | Linear | Constant | $k_s, k_v, k_0, v_0, d_0, t_h, \tau_a$ |
| | 42 | Linear contr. | CTH | None | Linear | MFC | $k_s, k_v, k_0, v_0, d_0, t_h, \tau_a$ |
| | 43 | Linear contr. | CTH | None | Nonlinear | None | $k_s, k_v, k_0, v_0, d_0, t_h, \tau_a, m_{load}, f_0, f_1, f_2$ |
| | 44 | Linear contr. | CTH | None | Nonlinear | Constant | $k_s, k_v, k_0, v_0, d_0, t_h, \tau_a, m_{load}, f_0, f_1, f_2$ |
| | 45 | Linear contr. | CTH | None | Nonlinear | MFC | $k_s, k_v, k_0, v_0, d_0, t_h, \tau_a, m_{load}, f_0, f_1, f_2$ |
| | 46 | Linear contr. | CTH | Constant | None | None | $k_s, k_v, k_0, v_0, d_0, t_h, \tau_p$ |
| | 47 | Linear contr. | CTH | Constant | None | Constant | $k_s, k_v, k_0, v_0, d_0, t_h, \tau_p$ |
| | 48 | Linear contr. | CTH | Constant | None | MFC | $k_s, k_v, k_0, v_0, d_0, t_h, \tau_p$ |
| | 49 | Linear contr. | CTH | Constant | Linear | None | $k_s, k_v, k_0, v_0, d_0, t_h, \tau_p, \tau_a$ |
| | 50 | Linear contr. | CTH | Constant | Linear | Constant | $k_s, k_v, k_0, v_0, d_0, t_h, \tau_p, \tau_a$ |
| | 51 | Linear contr. | CTH | Constant | Linear | MFC | $k_s, k_v, k_0, v_0, d_0, t_h, \tau_p, \tau_a$ |
| | 52 | Linear contr. | CTH | Constant | Nonlinear | None | $k_s, k_v, k_0, v_0, d_0, t_h, \tau_p, \tau_a, m_{load}, f_0, f_1, f_2$ |
| | 53 | Linear contr. | CTH | Constant | Nonlinear | Constant | $k_s, k_v, k_0, v_0, d_0, t_h, \tau_p, \tau_a, m_{load}, f_0, f_1, f_2$ |
| | 54 | Linear contr. | CTH | Constant | Nonlinear | MFC | $k_s, k_v, k_0, v_0, d_0, t_h, \tau_p, \tau_a, m_{load}, f_0, f_1, f_2$ |
| L-IDM-based models | 55 [a] | Linear contr. | IDM-desired | None | None | None | $k_s, k_v, k_0, v_0, d_0, t_h, a_{max}, a_{min}$ |
| | 56 | Linear contr. | IDM-desired | None | None | Constant | $k_s, k_v, k_0, v_0, d_0, t_h, a_{max}, a_{min}$ |
| | 57 | Linear contr. | IDM-desired | None | None | MFC | $k_s, k_v, k_0, v_0, d_0, t_h, a_{max}, a_{min}$ |
| | 58 | Linear contr. | IDM-desired | None | Linear | None | $k_s, k_v, k_0, v_0, d_0, t_h, a_{max}, a_{min}, \tau_a$ |
| | 59 | Linear contr. | IDM-desired | None | Linear | Constant | $k_s, k_v, k_0, v_0, d_0, t_h, a_{max}, a_{min}, \tau_a$ |
| | 60 | Linear contr. | IDM-desired | None | Linear | MFC | $k_s, k_v, k_0, v_0, d_0, t_h, a_{max}, a_{min}, \tau_a$ |
| | 61 | Linear contr. | IDM-desired | None | Nonlinear | None | $k_s, k_v, k_0, v_0, d_0, t_h, a_{max}, a_{min}, \tau_a, m_{load}, f_0, f_1, f_2$ |
| | 62 | Linear contr. | IDM-desired | None | Nonlinear | Constant | $k_s, k_v, k_0, v_0, d_0, t_h, a_{max}, a_{min}, \tau_a, m_{load}, f_0, f_1, f_2$ |
| | 63 | Linear contr. | IDM-desired | None | Nonlinear | MFC | $k_s, k_v, k_0, v_0, d_0, t_h, a_{max}, a_{min}, \tau_a, m_{load}, f_0, f_1, f_2$ |
| | 64 | Linear contr. | IDM-desired | Constant | None | None | $k_s, k_v, k_0, v_0, d_0, t_h, a_{max}, a_{min}, \tau_p$ |
| | 65 | Linear contr. | IDM-desired | Constant | None | Constant | $k_s, k_v, k_0, v_0, d_0, t_h, a_{max}, a_{min}, \tau_p$ |
| | 66 | Linear contr. | IDM-desired | Constant | None | MFC | $k_s, k_v, k_0, v_0, d_0, t_h, a_{max}, a_{min}, \tau_p$ |
| | 67 | Linear contr. | IDM-desired | Constant | Linear | None | $k_s, k_v, k_0, v_0, d_0, t_h, a_{max}, a_{min}, \tau_p, \tau_a$ |
| | 68 | Linear contr. | IDM-desired | Constant | Linear | Constant | $k_s, k_v, k_0, v_0, d_0, t_h, a_{max}, a_{min}, \tau_p, \tau_a$ |
| | 69 | Linear contr. | IDM-desired | Constant | Linear | MFC | $k_s, k_v, k_0, v_0, d_0, t_h, a_{max}, a_{min}, \tau_p, \tau_a$ |
| | 70 | Linear contr. | IDM-desired | Constant | Nonlinear | None | $k_s, k_v, k_0, v_0, d_0, t_h, a_{max}, a_{min}, \tau_p, \tau_a, m_{load}, f_0, f_1, f_2$ |
| | 71 | Linear contr. | IDM-desired | Constant | Nonlinear | Constant | $k_s, k_v, k_0, v_0, d_0, t_h, a_{max}, a_{min}, \tau_p, \tau_a, m_{load}, f_0, f_1, f_2$ |
| | 72 | Linear contr. | IDM-desired | Constant | Nonlinear | MFC | $k_s, k_v, k_0, v_0, d_0, t_h, a_{max}, a_{min}, \tau_p, \tau_a, m_{load}, f_0, f_1, f_2$ |
| L-Gipps-based models | 73 [a] | Linear contr. | Gipps-equilibrium | None | None | None | $k_s, k_v, k_0, v_0, d_0, t_h, \theta, a_{min}, \hat{a}_{min}$ |
| | 74 | Linear contr. | Gipps-equilibrium | None | None | Constant | $k_s, k_v, k_0, v_0, d_0, t_h, \theta, a_{min}, \hat{a}_{min}$ |
| | 75 | Linear contr. | Gipps-equilibrium | None | None | MFC | $k_s, k_v, k_0, v_0, d_0, t_h, \theta, a_{min}, \hat{a}_{min}$ |
| | 76 | Linear contr. | Gipps-equilibrium | None | Linear | None | $k_s, k_v, k_0, v_0, d_0, t_h, \theta, a_{min}, \hat{a}_{min}, \tau_a$ |
| | 77 | Linear contr. | Gipps-equilibrium | None | Linear | Constant | $k_s, k_v, k_0, v_0, d_0, t_h, \theta, a_{min}, \hat{a}_{min}, \tau_a$ |
| | 78 | Linear contr. | Gipps-equilibrium | None | Linear | MFC | $k_s, k_v, k_0, v_0, d_0, t_h, \theta, a_{min}, \hat{a}_{min}, \tau_a$ |
| | 79 | Linear contr. | Gipps-equilibrium | None | Nonlinear | None | $k_s, k_v, k_0, v_0, d_0, t_h, \theta, a_{min}, \hat{a}_{min}, \tau_a, m_{load}, f_0, f_1, f_2$ |
| | 80 | Linear contr. | Gipps-equilibrium | None | Nonlinear | Constant | $k_s, k_v, k_0, v_0, d_0, t_h, \theta, a_{min}, \hat{a}_{min}, \tau_a, m_{load}, f_0, f_1, f_2$ |
| | 81 | Linear contr. | Gipps-equilibrium | None | Nonlinear | MFC | $k_s, k_v, k_0, v_0, d_0, t_h, \theta, a_{min}, \hat{a}_{min}, \tau_a, m_{load}, f_0, f_1, f_2$ |
| | 82 | Linear contr. | Gipps-equilibrium | Constant | None | None | $k_s, k_v, k_0, v_0, d_0, t_h, \theta, a_{min}, \hat{a}_{min}, \tau_p$ |
| | 83 | Linear contr. | Gipps-equilibrium | Constant | None | Constant | $k_s, k_v, k_0, v_0, d_0, t_h, \theta, a_{min}, \hat{a}_{min}, \tau_p$ |
| | 84 | Linear contr. | Gipps-equilibrium | Constant | None | MFC | $k_s, k_v, k_0, v_0, d_0, t_h, \theta, a_{min}, \hat{a}_{min}, \tau_p$ |
| | 85 | Linear contr. | Gipps-equilibrium | Constant | Linear | None | $k_s, k_v, k_0, v_0, d_0, t_h, \theta, a_{min}, \hat{a}_{min}, \tau_p, \tau_a$ |
| | 86 | Linear contr. | Gipps-equilibrium | Constant | Linear | Constant | $k_s, k_v, k_0, v_0, d_0, t_h, \theta, a_{min}, \hat{a}_{min}, \tau_p, \tau_a$ |
| | 87 | Linear contr. | Gipps-equilibrium | Constant | Linear | MFC | $k_s, k_v, k_0, v_0, d_0, t_h, \theta, a_{min}, \hat{a}_{min}, \tau_p, \tau_a$ |
| | 88 | Linear contr. | Gipps-equilibrium | Constant | Nonlinear | None | $k_s, k_v, k_0, v_0, d_0, t_h, \theta, a_{min}, \hat{a}_{min}, \tau_p, \tau_a, m_{load}, f_0, f_1, f_2$ |
| | 89 | Linear contr. | Gipps-equilibrium | Constant | Nonlinear | Constant | $k_s, k_v, k_0, v_0, d_0, t_h, \theta, a_{min}, \hat{a}_{min}, \tau_p, \tau_a, m_{load}, f_0, f_1, f_2$ |
| | 90 | Linear contr. | Gipps-equilibrium | Constant | Nonlinear | MFC | $k_s, k_v, k_0, v_0, d_0, t_h, \theta, a_{min}, \hat{a}_{min}, \tau_p, \tau_a, m_{load}, f_0, f_1, f_2$ |

[a] Five base models (without physical extensions). ***Acronyms***: ACC = adaptive cruise control; IDM = intelligent driver model; CTH = constant-time-headway; MFC = microsimulation free-flow acceleration boundary model. ***Symbols***: $\delta$ = exponent factor of IDM; $v_0$ = ACC set speed (m/s); $d_0$ = minimum desired spacing (m); $t_h$ = time headway (s); $a_{max}$ = maximum acceleration (m/s$^2$); $a_{min}$ = comfort deceleration rate (m/s$^2$, negative); $\tau_p$ = perception delay (s); $\tau_a$ = actuation lag (s); $m_{load}$ = vehicle load mass (kg); $f_0$, $f_1$, and $f_2$ = road load coefficients (N, kg/s, and kg/m, respectively); $\theta$ = delay factor (s) in Gipps' model; $\hat{a}_{min}$ = estimated preceding vehicle deceleration rate (m/s$^2$, negative); $k_s$ = gain factor (s$^{-2}$) for the spacing error; $k_v$ = gain factor (s$^{-1}$) for the speed error; $k_0$ = gain factor (s$^{-1}$) for speed deviation from the desired speed.

# 3. Field platoon data

Vehicle trajectory data used in this study were collected in platooning experiments carried out on the Rural Road of the AstaZero proving ground (Sweden) in 2019. The test track is 5.7-km long, with slopes ranging from -3 to 3%. Fig. 2 shows the alignment of the track layout.

The tested platoon was composed by five high-end vehicles from different manufacturers (Audi, Tesla, Mercedes, and BMW), all equipped with commercial ACC. Trajectory data were acquired with the RT-Range S multiple target ADAS measurements solution, supplied by Oxford Technical Solutions Company. The system has a sampling rate of 100 Hz (trajectory data were down sampled to 10 Hz), and guarantees precisions of 0.02 m/s and 0.02 m for speed and position measurements, respectively (Makridis et al., 2021).

Table 2 lists the vehicle compositions (platoon leader and four followers) in 7 platoons (some platoons shared the same composition). Data analysis showed that the platoon was in car-following regime during the whole experiment, i.e., the follower vehicle dynamics were influenced more by the preceding vehicle trajectory, vehicle characteristics and onboard ACC system setup, than by road geometry or slope.

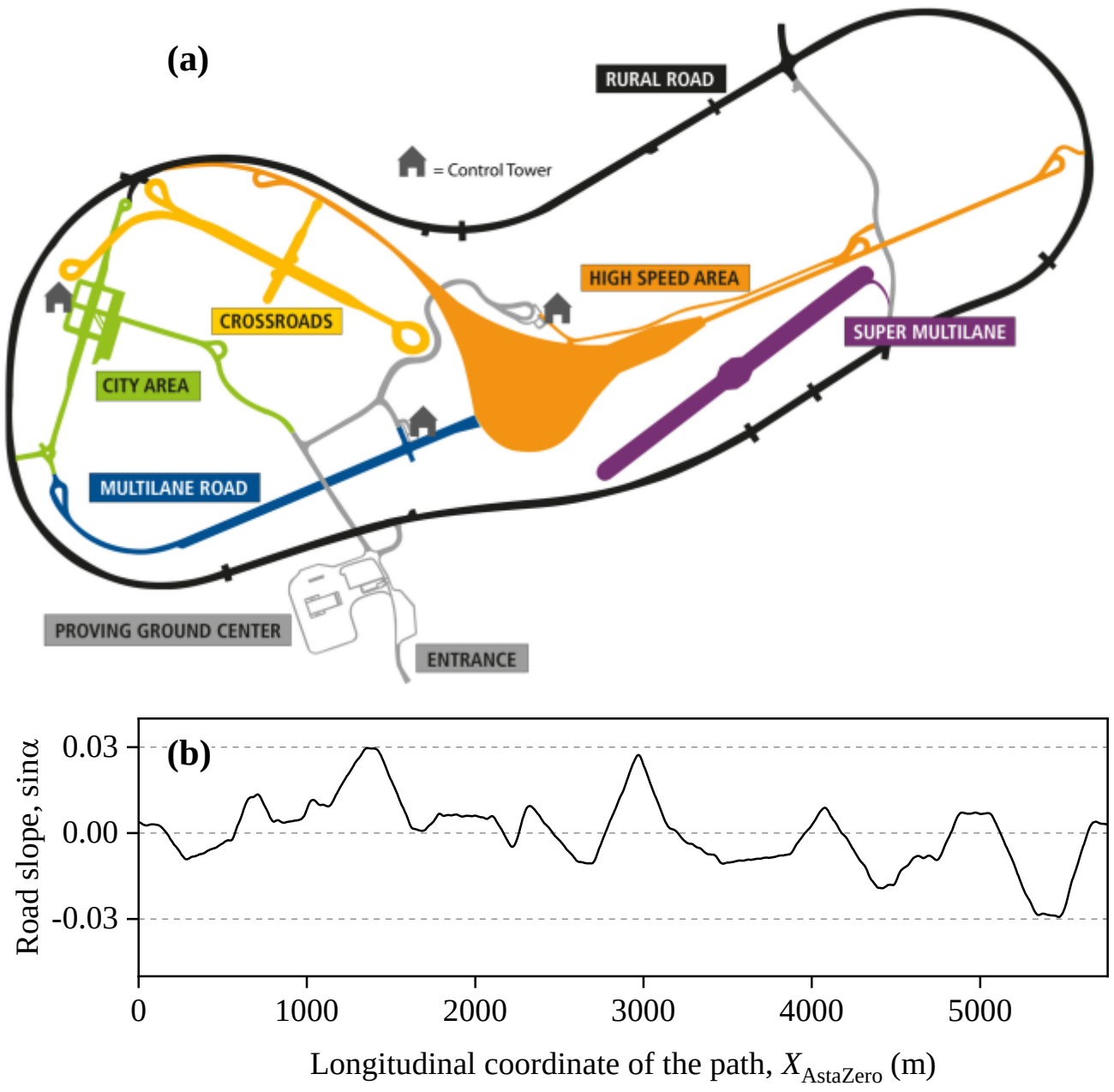

**Fig. 2. AstaZero proving ground in Sweden: (a) path layout and (b) road slope. ([https://www.astazero.com](https://www.astazero.com))**

**Table 2. Composition of platoons.**

| Vehicle position | Platoons P1 | P2 | P3 | P4 | P5 | P6 | P7 |
|---|---|---|---|---|---|---|---|
| Leader | Audi A8 | Audi A8 | Audi A8 | Audi A8 | Audi A8 | Audi A8 | Audi A8 |
| Follower 1 | Audi A6 | Audi A6 | Tesla | Tesla | Tesla | Mercedes | Mercedes |
| Follower 2 | BMW | BMW | BMW | BMW | BMW | BMW | BMW |
| Follower 3 | Mercedes | Mercedes | Audi A6 | Audi A6 | Audi A6 | Tesla | Tesla |
| Follower 4 | Tesla | Tesla | Mercedes | Mercedes | Mercedes | Audi A6 | Audi A6 |

Fig. 3 illustrates vehicle trajectory data of 7 platoons tested on the AstaZero proving ground. Fig. 3 (a)-(g) provide the speed profiles of the platoon leader and four following vehicles. Fig. 3 (h)-(n) give the corresponding spacing profiles. In the tests, the platoon leader (always an Audi A8), was instructed to keep a predefined speed value (via the onboard ACC system), in order to prevent speed fluctuations caused by human driving. Speed perturbations were then introduced in the platoon by changing the set speed ($v_0$) of the leading vehicle's ACC system. For safety purposes, the ACC set speed lied between 13.9 and 16.7 m/s in curves, and between 25 and 27.8 m/s in straight sections.

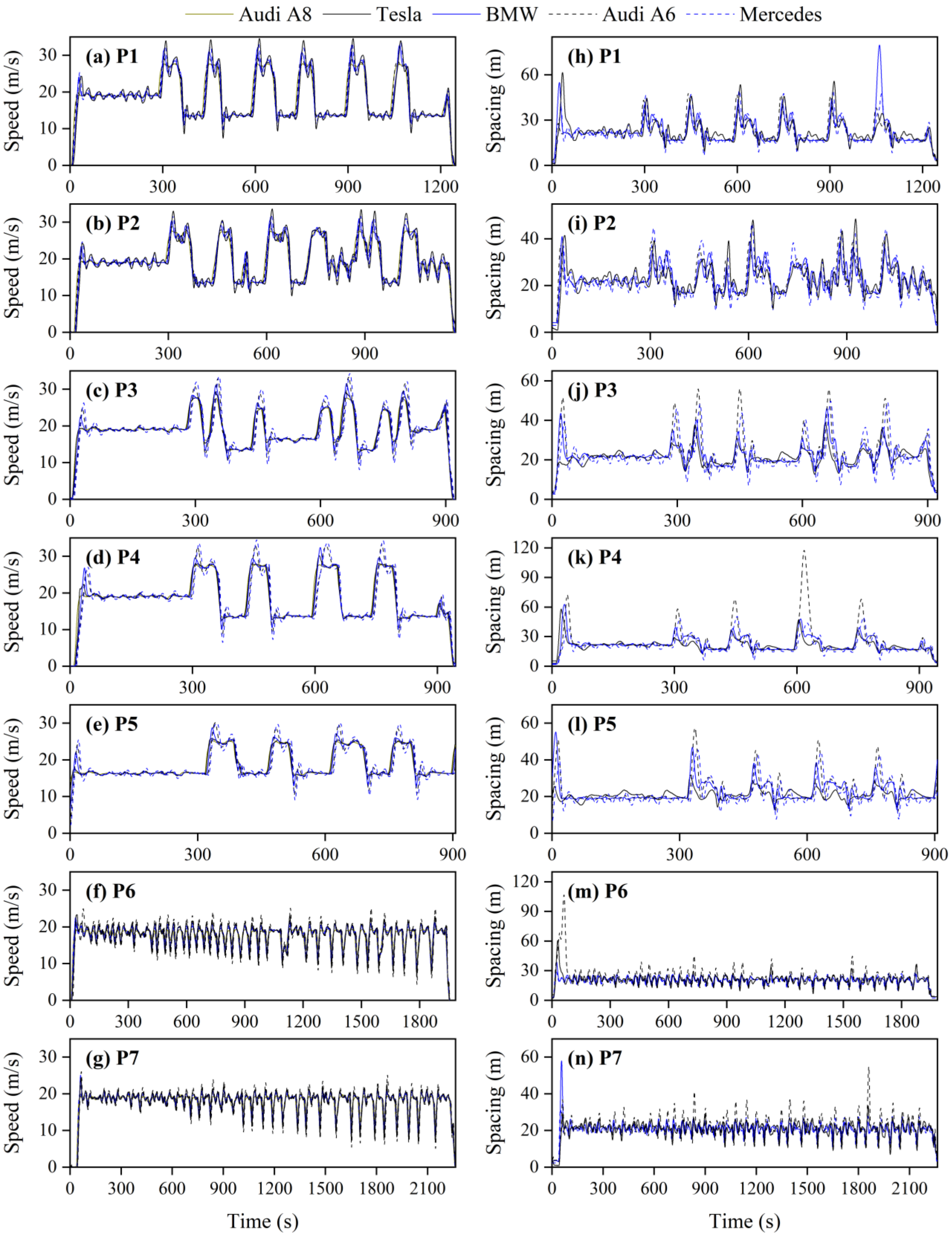

**Fig. 3. Field data of seven platoons tested on the AstaZero proving ground: (a)-(g) speed (m/s) and (h)-(n) spacing (m).**

# 4. Methodology and Design of Experiments

In order to appraise the impact of increasingly complex lower-level dynamics on model ability to reproduce observed ACC vehicle dynamics, models listed in Table 1 have been calibrated against ACC trajectory data. By comparing calibration errors among the variants of each base model, the contribution of every physics-enhancement to model accuracy has been outlined.

In addition, by comparing variants of different base models among each other, it has been also possible to verify which model is more likely to reproduce observed ACC vehicle dynamics, and if the consolidated habit of applying linear control laws for this purpose, instead of CF models, has practical significance.

To investigate the robustness of calibrated models in terms of transferability of model parameters, cross-validation experiments have been also conducted. In these experiments, the vehicle trajectory in a platoon has been simulated using the set of parameter values obtained by calibrating the model against the trajectory of the same vehicle in a different platoon. The cross-validation analysis has also revealed which model is more likely to avoid vehicle collisions, i.e., generation of negative inter-vehicle spacings, which is the effect of string instability (Montanino and Punzo, 2021; Montanino et al., 2021).

## 4.1. Methodology

A general formulation of the model calibration problem is described as follows (Punzo et al., 2021):

$$\begin{gathered} MoP^{sim} = F(\boldsymbol{\beta}) \\ minimize\ f(MoP^{obs}, MoP^{sim}) \\ subject\ to\colon LB_{\boldsymbol{\beta}} \leq \boldsymbol{\beta} \leq UB_{\boldsymbol{\beta}}, \boldsymbol{G}(\boldsymbol{\beta}) \leq \mathbf{0} \end{gathered} \tag{17}$$

where $\boldsymbol{\beta}$ is a vector of model parameters to be calibrated (the number of parameters can change depending on the model variant); $F(\cdot)$ is a CF/ACC model variant, that is a function of $\boldsymbol{\beta}$; $LB_{\boldsymbol{\beta}}$ and $UB_{\boldsymbol{\beta}}$ represent the lower and the upper bound for the parameters in $\boldsymbol{\beta}$, respectively; $\boldsymbol{G}(\cdot)$ is a vector of constraint functions; $\boldsymbol{MoP^{obs}}$ and $\boldsymbol{MoP^{sim}}$ represent observed and simulated Measure of Performance, respectively; $\boldsymbol{f}(\cdot)$ is a goodness-of-fit (*GoF*) function; and $f(MoP^{obs}, MoP^{sim})$ is the optimization objective function. The aim of CF model calibration is to minimize $f(MoP^{obs}, MoP^{sim})$.

Basing on the findings in Punzo et al. (2021), the Normalized Root Mean Square Error of spacing, speed, and acceleration, i.e., *NRMSE*(*s*,*v*,*a*), has been adopted as *GoF* function in model calibration. In fact, in that paper, *NRMSE*(*s*,*v*,*a*) was proved to be the most preferable objective function to be used in a CF calibration problem, basing on theoretical arguments and wide empirical evidence. The formulation of *NRMSE*(*s*,*v*,*a*) is the following:

$$GoF\colon\ f = \mathrm{NRMSE}(Y)$$

$$MoP\colon\ Y = \{s, v, a\}$$

$$\begin{cases} \mathrm{NRMSE}(s,v,a) = \beta_0 \mathrm{NRMSE}(s) + \beta_1 \mathrm{NRMSE}(v) + \beta_2 \mathrm{NRMSE}(a) \\ \mathrm{NRMSE}(Y) = \mathrm{RMSE}(Y) \Big/ \sqrt{\frac{1}{T}\sum_{\tau_p+1}^{T}\left(Y^{\mathrm{obs}}(t)\right)^2} \\ \mathrm{RMSE}(Y) = \sqrt{\frac{1}{T}\sum_{\tau_p+1}^{T}\left(Y^{\mathrm{sim}}(t) - Y^{\mathrm{obs}}(t)\right)^2} \end{cases} \tag{18}$$

where $\beta_0$, $\beta_1$, and $\beta_2$ are weight factors, which are assumed equal to 1 in this study (i.e., model accuracy on each *MoP* is equally desirable); $T$ is the total number of measurement instants in an experimental trajectory; $\tau_p$ is the model perception delay.

In order to assess the ability of the whole calibration setting (optimization algorithm and *GoF*(*MoPs*) function) to find robust solutions in terms of *GoF* value, 10 replicates of each model calibration experiment have been performed. By comparing *GoF* value among replicates, we have verified whether the algorithm was able to converge to the same minimum at each calibration replication. The optimization algorithm used in this study is the Genetic Algorithm coded in Python (https://pypi.org/project/geneticalgorithm/).

Eventually, to test the robustness of the chosen calibration *GoF*, against alternative *GoF*s, models have been calibrated also by means of the following *GoF*s: *RMSE*($s$), *RMSE*($v$), Theil's $U(s,v)$, Theil's $U(s,v,a)$, and *NRMSE*($s,v$) (for their definitions, see Punzo et al., 2021).

Finally, to investigate the accuracy of calibrated models in validation, performances have been evaluated via the *NRMSE*($s,v,a$).

## 4.2. Design of experiments

The calibration and validation experiments have been conducted for the 90 models listed in Table 1, against trajectory data of 4 following vehicles, in 7 platoons.

To allow a better understanding of the design of experiments, an index of ($Veh$, $P_{cal}$, $P_{val}$) is introduced to represent both calibration and validation experiments, where $Veh \in$ {Tesla, BMW, Audi A6, Mercedes} is the vehicle type; $P_{cal}$ and $P_{val} \in$ {P1, P2, P3, P4, P5, P6, P7} denote the platoon ID used, respectively, for calibration and validation. For example, the experiment with index (Tesla, P2, P6) means that model parameters have been calibrated against the trajectory of the vehicle Tesla in the platoon P2, and that resulting optimal parameters have been adopted to simulate the trajectory of the same vehicle in platoon P6.

Therefore, experiments with $P_{cal} = P_{val}$ denote *calibration* experiments, which are 28 per model, i.e., 4 vehicles × 7 platoons. A total of 2520 calibration experiments (28 × 90 models) have been performed.

Experiments with $P_{cal} \neq P_{val}$ refer to *validation* experiments, which are 168 per model, i.e., 28 sets of calibrated parameters (4 vehicles × 7 platoons) × 6 simulations (where 6 is the number of platoons in which the same vehicle is observed, other than $P_{cal}$).

Each simulation has been performed following the guidelines provided in Punzo et al. (2021), i.e.,

- Time step consistency: the time step of car-following model simulation is consistent with the resolution of trajectory data (it is assumed equal to 0.1 s in this work).
- Data internal consistency: since available data consisted in GPS positions, ego vehicle and preceding vehicle speeds, as well as accelerations, have been derived using the same integration scheme applied for the model numerical integration (see Eq. 1).

Table 3 lists the lower and the upper bounds of all calibration parameters, shared among the 90 models (for parameters definition, see also Table 1). Since road load coefficients ($f_0$, $f_1$, and $f_2$) vary with the vehicle type, their bounds are written as a four-dimensional vector (each dimension is a vehicle type). The medians of these ranges have been estimated on the chassis dynamometer so that the ranges have been intentionally kept small around the estimated values (wider ranges could cause a compensation of the model uncertainty and thus result in overfitting in calibration; see Punzo and Montanino, 2020).

**Table 3. Bounds of calibration parameters.**

| Parameter [unit] | Lower bound | Upper bound |
|---|---|---|
| $\delta$ | 0.1 | 10 |
| $v_0$ [m/s] | 30 | 35 |
| $d_0$ [m] | 1 | 5 |
| $t_h$ [s] | 0.1 | 3 |
| $a_{max}$ [m/s$^2$] | 0.5 | 5 |
| $a_{min}$ [m/s$^2$] | -5 | -0.5 |
| $\hat{a}_{min}$ [m/s$^2$] | -5 | -0.5 |
| $\theta$ [s] | 0 | 3 |
| $\tau_a$ [s] | 0.3 | 0.8 |
| $\tau_p$ [s] | 0.1 | 0.8 |
| $m_{load}$ [kg] | 230 | 300 |
| $f_0$ [N] | [185.1, 190.3, 154.9, 139.5] [a] | [226.3, 232.6, 189.4, 170.4] [a] |
| $f_1$ [kg/s] | [0, -0.63, 0.64, 0.56] [a] | [0, -0.52, 0.78, 0.69] [a] |
| $f_2$ [kg/m] | [0.025, 0.042, 0.026, 0.027] [a] | [0.031, 0.051, 0.031, 0.033] [a] |
| $k_s$ [s$^{-2}$] | 0.01 | 5 |
| $k_v$ [s$^{-1}$] | 0.01 | 5 |
| $k_0$ [s$^{-1}$] | 0.01 | 5 |

[a] Bounds for different vehicle types (Tesla, BMW, Audi A6, and Mercedes, respectively).

# 5. Calibration results

## 5.1. Robustness of the calibration setting

In CF model calibration, the strong nonlinearity of the model response surface, the large-dimensional solution space and the stochasticity of certain optimization algorithms, raise the possibility of trapping the algorithm in a local optimum, and of causing poor results reproducibility.

To address this concern, we have run 10 replicates of each calibration experiment, and we have adopted the coefficient of variation (CV) of obtained *NRMSE*(*s*,*v*,*a*) values at optimum as a measure of relative variability across the replicates. The analysis of CV distribution of all calibration experiments (i.e., 2520 calibration experiments) allowed us to assess the ability of a calibration setting to find robust solutions in terms of *GoF* value. In fact, if the CV distribution result highly concentrated at the zero-value, i.e., if standard deviation of *GoF* values among replicas is very close to 0 in each experiment, calibration results do not change when running the same experiment multiple times.

CV distributions are shown in Fig. 4, which show CV distributions grouped by model class. In each subplot, the distribution of 504 CV values is presented, each value measuring the variation across 10 replicates of each of the 504 calibration experiments (18 models × 4 vehicle types × 7 platoon trajectories).

Focusing on the CV variability range, findings suggest that *GoF* values at optimum, for all five model classes, are highly reproducible (CVs are always less than 3%). To infer on zero-value concentration, data have been also fitted by an exponential distribution with a decay rate parameter $\lambda$, which measures how rapidly the density declines as the CV value increases. IDM-based models stand out prominently in terms of calibration results reproducibility, showing a $\lambda$ value which is almost one order of magnitude higher than the $\lambda$ values of the other model classes.

Concerning the choice of the objective function, we have verified that model calibrations using the *NRMSE*(*s*,*v*,*a*) have returned the lowest sum of relative errors on spacing, speed and acceleration, compared to calibrations experiments with other objective functions, i.e. *RMSE*(*s*), *RMSE*(*v*), Theil's *U*(*s*,*v*), Theil's *U*(*s*,*v*,*a*), and *NRMSE*(*s*,*v*). The relative error measures the degradation in a *MoP* – relative to its optimum – of a model calibrated using a certain objective function. For more details on the methodology to compare objective functions, the reader may refer to Punzo et al. (2021).

This result confirms the goodness of the recommendation provided in Punzo et al. (2021) about preferable calibration settings, and extends the validity of the recommendation to a considerably larger number of models (90 in this study) and to a larger dataset of ACC trajectories.

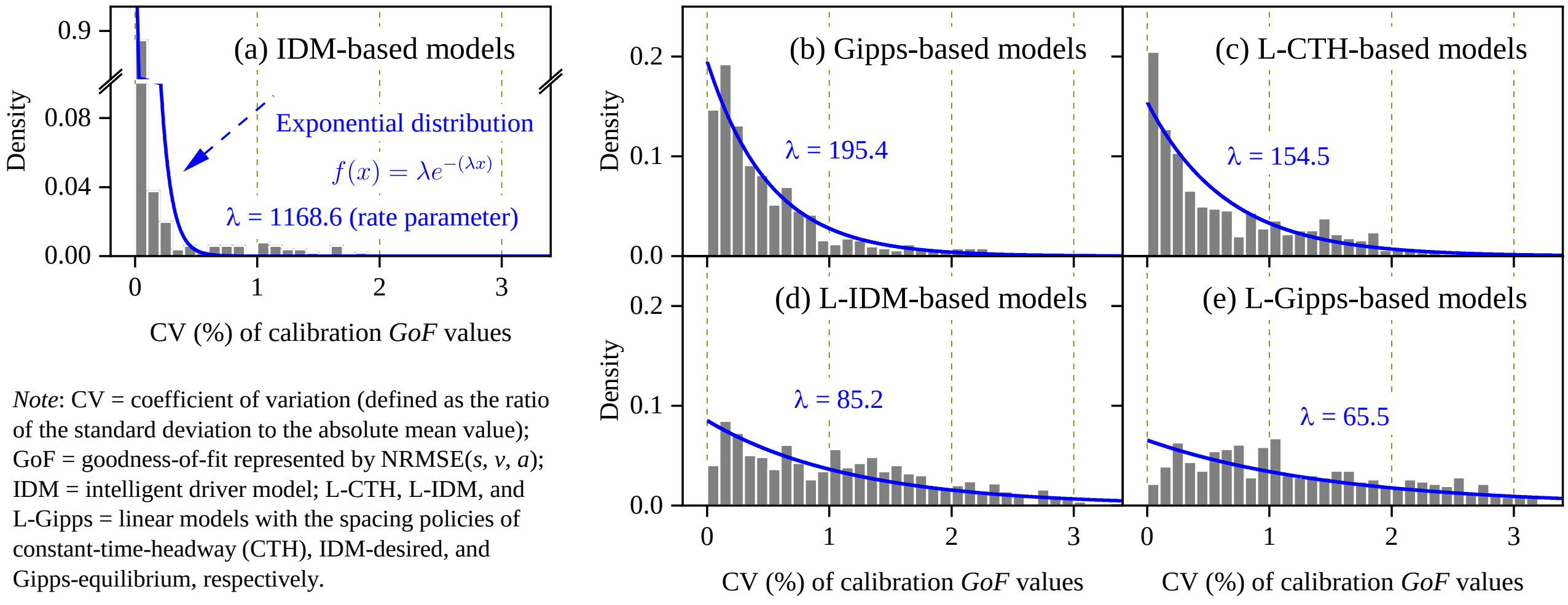

*Note*: CV = coefficient of variation (defined as the ratio of the standard deviation to the absolute mean value); GoF = goodness-of-fit represented by NRMSE(*s*, *v*, *a*); IDM = intelligent driver model; L-CTH, L-IDM, and L-Gipps = linear models with the spacing policies of constant-time-headway (CTH), IDM-desired, and Gipps-equilibrium, respectively.

**Fig. 4. Distribution of the coefficient of variation (CV) of *NRMSE*(*s*,*v*,*a*): (a) IDM-based models, (b) Gipps-based models, (c) L-CTH-based models, (d) L-IDM-based models, and (e) L-Gipps-based models**

## 5.2. Comparison of base models across vehicles and platoons

Fig. 5 compares the errors of the five base models after calibration against the trajectories of the 4 vehicles in each of the 7 platoons. Each sub-plot shows errors variability across the 4 vehicles in a specific platoon (sorted according to Table 2). For a given vehicle, the first and the last bar (white and gold bars), show the minimum and maximum *NRMSE*(*s*,*v*,*a*) values achieved after calibration among all ninety models. They are the best/worst values that can be achieved with either base models or their variants. The five bars in the middle of each group tell the *NRMSE* obtained from the calibration of the five base models.

The ability to reproduce observed ACC dynamics vary more with the trajectory data (vehicle and platoon) than with the model type. In fact, the error variance across base models is lower than the variance across vehicles. In addition, given a vehicle, model performances can sensibly vary depending on the platoon, i.e., on the input leader trajectory and on the vehicle order in the platoon (see, for instance, the Audi A6 across the 7 platoons, and the high calibration errors obtained in P4 and P6, which are mainly driven by the spacing peaks shown in Fig. 3(k) and (m)). These results are in accordance with those in the literature of car-following models (see e.g. Punzo and Simonelli, 2005).

Results revealed that, despite the absence of time stochasticity of ACC trajectories, no base model was able to perform consistently on a given vehicle type, regardless of platoon leader dynamics and vehicle order in the platoon.

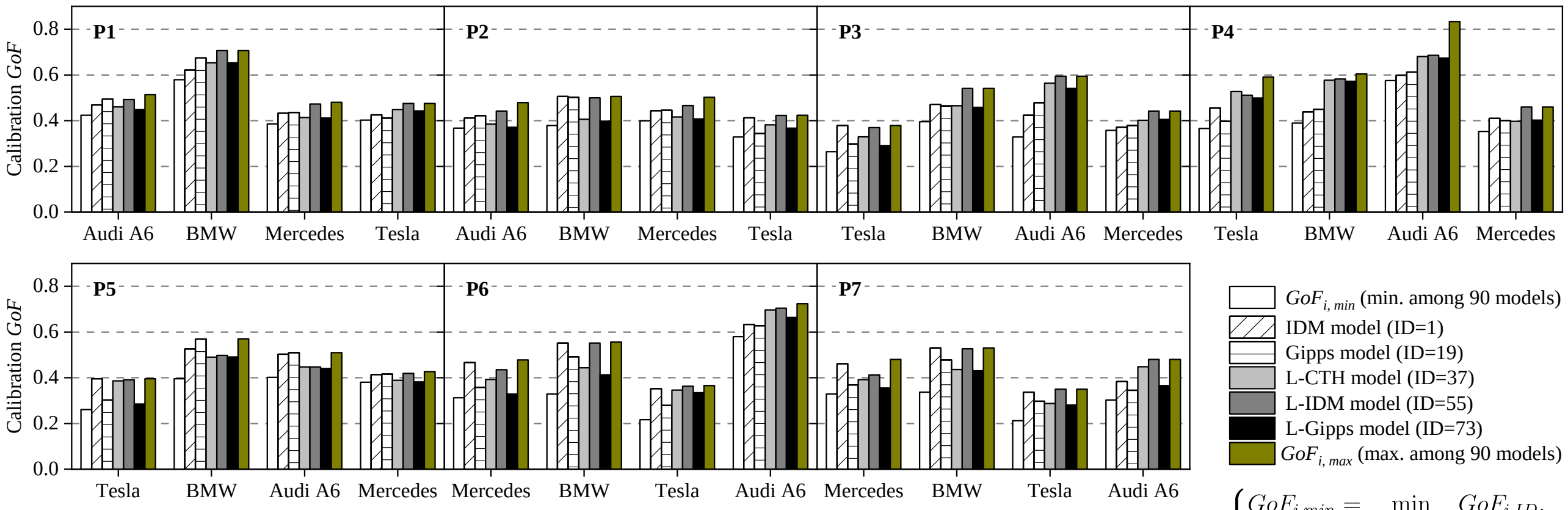

*Note*: GoF = goodness-of-fit represented by NRMSE(*s*, *v*, *a*); IDM = intelligent driver model; L-CTH, L-IDM, and L-Gipps = linear models with the spacing policies of constant-time-headway (CTH), IDM-desired, and Gipps-equilibrium, respectively.

$$\begin{cases} GoF_{i,min} = \min\limits_{ID=1,\ldots,90} GoF_{i,ID}, \\ GoF_{i,max} = \max\limits_{ID=1,\ldots,90} GoF_{i,ID}. \end{cases}$$

**Fig. 5. Variability of calibration errors of the five base models, across the 28 vehicle trajectories. For the *i*-th trajectory, the calibration *GoF* values of the five base models (*ID* = 1, 19, 37, 55, and 73 in Table 1), as well as the minimum and maximum *GoF* values among all 90 models ($GoF_{i,min}$ and $GoF_{i,max}$), are shown.**

## 5.3. Comparison of model variants

Fig. 6 shows a box-whisker plot of normalized calibration errors of the ninety model variants (i.e., $\widetilde{GoF}_{i,ID}$).

To compute the boxplot of each model variant, 28 calibration experiments have been run (i.e., one for each of the 28 trajectories). The objective function value of a calibrated model in the *i*-th calibration experiment – i.e. the calibration error of model *ID* for trajectory *i* – has been indicated by $GoF_{i,ID}$. Given a trajectory *i* and a model *ID*, the calibration error $GoF_{i,ID}$ has been *normalized* over the minimum calibration error for trajectory *i*, among all 90 models:

$$\widetilde{GoF}_{i,ID} = \frac{GoF_{i,ID} - GoF_{i,min}}{GoF_{i,min}}, \quad \text{with: } \begin{cases} i \in [1, \ldots, 28], \\ ID \in [1, \ldots, 90], \\ GoF_{i,min} = \min\limits_{ID=1,\ldots,90} GoF_{i,ID}, \end{cases} \tag{19}$$

In this way, we can rule out the direct effect of trajectory data variability, which results in very different calibration errors among trajectories, and which would make impossible a model comparison. In addition, in model comparison, we do not care about the absolute performance of a model, but we are interested in its performance relative to those of other models in the same calibration experiment/trajectory.

In Fig. 6, the horizontal line and the square within each box indicate the median and mean values of normalized calibration errors. The bottom and top edges of the box represent the 1st (Q1) and 3rd (Q3) quartiles, respectively. The maximum whisker lengths correspond to 1.5 times the interquartile range (IQR). Outliers, i.e., calibration experiments returning a normalized error greater than 1.5IQR plus the third quartile, are marked with black diamonds.

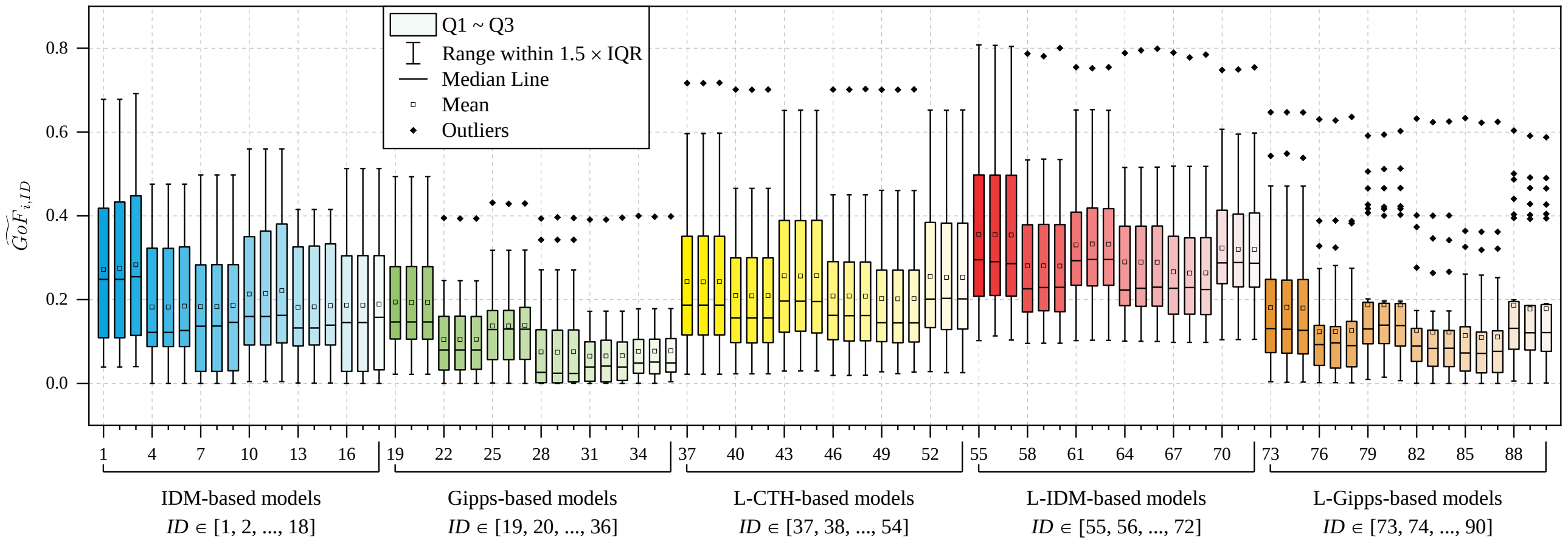

*Note*: $GoF_i$ = goodness-of-fit function value for the *i*-th trajectory, calculated by NRMSE(*s*, *v*, *a*); IDM = intelligent driver model; L-CTH, L-IDM, and L-Gipps = linear models with the spacing policies of constant-time-headway (CTH), IDM-desired, and Gipps-equilibrium, respectively; IQR = interquartile range; Q = quartile.

$$\widetilde{GoF}_{i,ID} = \frac{GoF_{i,ID} - GoF_{i,min}}{GoF_{i,min}} \text{ and } GoF_{i,min} = \min_{ID=1,\dots,90} GoF_{i,ID},$$
$$\text{for } i \in [1, \dots, 28] \text{ and } ID \in [1, \dots, 90]$$

**Fig. 6. Variability of normalized calibration errors within each model (individual box plot) and among models. Each box plot represents the distribution of 28 calibration errors. The *i*-th $GoF_{i,ID}$ calibration error value of the *ID*-th model is normalized over the corresponding minimum error value among all models, $GoF_{i,min}$.**

Boxplots are grouped into five base model classes and are coloured accordingly.

IDM-based models contain no outliers, suggesting more robust calibration performances.

Gipps-based and L-Gipps-based models show the best calibration performances (L-Gipps-based models are less robust than Gipps-based ones, as to the number of outliers). We recall here that the Gipps' safety-based assumption – i.e., the following vehicle maintains a speed which ensures a safe stop even in case of a sudden and severe deceleration, up to a stop, of the preceding vehicle – is inherent in Gipps-based models, and is transferred into L-Gipps-based models through Gipps' equilibrium speed-distance function, which is adopted there as spacing policy.

If we focus on linear controllers, L-Gipps-based models perform better than linear controllers with an IDM desired-distance spacing policy or with a CTH spacing policy (which is the algorithm customarily applied in the field literature to study ACCs).

Concerning the impact of physics-extensions on model accuracy, a useful comparison can be made by means of the median values of the errors in Figure 6 (see the horizontal lines in each box plot). The figure reveals that linear vehicle dynamics largely improve (median) model performances of all base models (see the second triple of boxes in each model class, which is always shifted down relative to the first triple). In turn, nonlinear vehicle dynamics are not as beneficial as linear ones, and, in some cases, performances deteriorate relative to base models (an explanation of this effect is provided below).

Perception delay also sensibly improves modelling accuracy of all base models. Such increase has the same order of magnitude of that produced by linear dynamics. A possible explanation is that the actuation lag in linear dynamics, and the perception delay factor, produce similar delayed vehicle dynamics in simulation.

However, the combined effect of linear dynamics and perception delay is less than the sum of the single contributions (i.e., adding both is not beneficial to model accuracy).

To quantify the effect of physics-enhancements among model variants, Table 4 reports the median normalized errors from Fig. 6, together with the percentage variation from the base models. The table allows us to investigate both the individual and the combined effect of physics-extensions on model accuracy. In the first column, rows are grouped by main physics-extensions applied to the base model, which are common to all variants in the group.

Among the extensions, the impact of acceleration constraints on model accuracy is lower than the impact produced by vehicle dynamics and perception delay. Moreover, since the acceleration constraints limit the space of feasible optimal solutions – in fact in absence of constraints the model logic does not take into account e.g. motor limits and, therefore, the returned acceleration command can violate such physical limits – the accuracy of model variants with constraints (i.e., ID 2, 3, 5, 6, 8, 9, 11, 12, 14, 15, 17, 18 in Table 4) is more likely to be lower than that of models without constraints (i.e., ID 1, 4, 7, 10, 13, 16; see also Remark in Section 2.2; few exceptions, however, occurred, when the calibrated model without constraints returned aggressive behaviours). The same reasoning applies also to nonlinear vehicle dynamics variants, which show lower accuracy than linear ones (and sometimes even lower accuracy than the base model). In fact, being the calibration ranges of road load coefficients very tight (see motivation in Section 4.2), the space of feasible solutions for model variants with nonlinear dynamics is more limited than the one for linear dynamics variants, thus the calibrated models return higher errors.

**Table 4.** Median and percentage variation (relative to the base model) of normalized calibration errors.

| | IDM-based models | | | Gipps-based models | | | L-CTH-based models | | | L-IDM-based models | | | L-Gipps-based models | | |
|---|---|---|---|---|---|---|---|---|---|---|---|---|---|---|---|
| Main extensions | ID | $\widetilde{GoF}_{i,ID}$ median | ($\widetilde{GoF}_{i,ID}$-base)/base, % | ID | $\widetilde{GoF}_{i,ID}$ median | ($\widetilde{GoF}_{i,ID}$-base)/base, % | ID | $\widetilde{GoF}_{i,ID}$ median | ($\widetilde{GoF}_{i,ID}$-base)/base, % | ID | $\widetilde{GoF}_{i,ID}$ median | ($\widetilde{GoF}_{i,ID}$-base)/base, % | ID | $\widetilde{GoF}_{i,ID}$ median | ($\widetilde{GoF}_{i,ID}$-base)/base, % |
| | 1 [a] | 0.248 | - | 19 [a] | 0.147 | - | 37 [a] | 0.187 | - | 55 [a] | 0.295 | - | 73 [a] | 0.131 | - |
| | 2 | 0.248 | 0 | 20 | 0.147 | 0 | 38 | 0.187 | 0 | 56 | 0.291 | -2 | 74 | 0.130 | -1 |
| | 3 | 0.255 | 3 | 21 | 0.147 | 0 | 39 | 0.187 | 0 | 57 | 0.286 | -3 | 75 | 0.127 | -3 |
| Linear veh. dyn. (LVD) | 4 | 0.122 | -51 | 22 | 0.080 | -46 | 40 | 0.157 | -16 | 58 | 0.226 | -24 | 76 | 0.093 | -29 |
| | 5 | 0.122 | -51 | 23 | 0.080 | -45 | 41 | 0.157 | -16 | 59 | 0.229 | -23 | 77 | 0.097 | -26 |
| | 6 | 0.126 | -49 | 24 | 0.080 | -45 | 42 | 0.157 | -16 | 60 | 0.230 | -22 | 78 | 0.091 | -31 |
| Nonlinear veh. dyn. (NLVD) | 7 | 0.137 | -45 | 25 | 0.129 | -12 | 43 | 0.197 | 5 | 61 | 0.293 | -1 | 79 | 0.130 | 0 |
| | 8 | 0.137 | -45 | 26 | 0.130 | -11 | 44 | 0.196 | 5 | 62 | 0.296 | 0 | 80 | 0.139 | 6 |
| | 9 | 0.146 | -41 | 27 | 0.129 | -12 | 45 | 0.195 | 4 | 63 | 0.296 | 0 | 81 | 0.139 | 6 |
| Percept. delay (PD) | 10 | 0.160 | -36 | 28 | 0.027 | -82 | 46 | 0.162 | -13 | 64 | 0.223 | -25 | 82 | 0.089 | -32 |
| | 11 | 0.160 | -36 | 29 | 0.025 | -83 | 47 | 0.162 | -14 | 65 | 0.227 | -23 | 83 | 0.084 | -36 |
| | 12 | 0.163 | -34 | 30 | 0.024 | -84 | 48 | 0.162 | -13 | 66 | 0.230 | -22 | 84 | 0.084 | -36 |
| PD + LVD | 13 | 0.132 | -47 | 31 | 0.039 | -73 | 49 | 0.145 | -22 | 67 | 0.227 | -23 | 85 | 0.073 | -45 |
| | 14 | 0.132 | -47 | 32 | 0.042 | -71 | 50 | 0.145 | -23 | 68 | 0.229 | -23 | 86 | 0.072 | -45 |
| | 15 | 0.139 | -44 | 33 | 0.039 | -73 | 51 | 0.145 | -23 | 69 | 0.224 | -24 | 87 | 0.077 | -42 |
| PD + NLVD | 16 | 0.146 | -41 | 34 | 0.049 | -67 | 52 | 0.202 | 8 | 70 | 0.288 | -3 | 88 | 0.132 | 0 |
| | 17 | 0.146 | -41 | 35 | 0.051 | -65 | 53 | 0.203 | 9 | 71 | 0.289 | -2 | 89 | 0.121 | -8 |
| | 18 | 0.158 | -36 | 36 | 0.049 | -66 | 54 | 0.202 | 8 | 72 | 0.287 | -3 | 90 | 0.121 | -7 |

[a] Base model.

# 6. Validation results

## 6.1. Robustness to vehicle collision

Model validation experiments consist in simulating a trajectory with a model calibrated on a different trajectory. When models calibrated on a trajectory are used to simulate a different trajectory, they can yield a collision, i.e., a negative inter-vehicle spacing.[1] Since observed vehicle dynamics in all platoons are collision-free, the emergence of a collision in simulation is an evidence that calibrated parameters are string unstable and that, though optimal for the specific trajectory used in calibration, they cannot be transferred to simulate different trajectories.

Therefore, the frequency of collisions measured in validation experiments is an index of model robustness and parameter transferability (Zhu et al., 2018). Fig. 7 shows the collision frequency of each model over 168 validation experiments (4 vehicles × 7 platoons × 6 simulations/trajectories; see Section 4.2). In order to evaluate also the impact of different calibration settings on parameter robustness and transferability, we have plotted six curves, each one corresponding to a calibration performed with a different *GoF*(*MoP*).

[1] In model calibration, if a vehicle collision occurs, the *GoF* value is typically set to a very large number, in order to exclude the corresponding tuple of parameter values from the space of feasible solutions (this approach is customary in the literature and is equivalent to adding a nonlinear constraint in the optimization problem; moreover, it avoids performing an extra model simulation, which is necessary to evaluate the nonlinear constraint). Consequently, the solution returned by the optimization algorithm cannot produce vehicle collisions when the model is simulated in the same car-following experiment adopted for model calibration.

*NRMSE*(*s*,*v*,*a*) and Theil's *U*(*s*,*v*,*a*) have provided the least number of collisions, for all ninety models.

On the one hand, this result extends the validity of the recommendations given in Punzo et al. (2021). In that paper, it has been shown that these two *GoFs* are the most preferable for CF model calibration in terms of accuracy, among the Pareto efficient ones (they also outperform objective functions with a single *MoP*). Figure 7 shows that the two mentioned *GoFs* guarantee also the highest robustness and transferability of calibrated parameters.

On the other hand, when a single *MoP* is applied, plotted results show that, also in terms of robustness and transferability of calibrated parameters, spacing is preferable to speed (concerning the dualism between spacing and speed in calibration of car-following models see Punzo and Montanino, 2016). In fact, the *RMSE*(*v*) displays the worst performances, since it sensibly increases the collision frequency in L-CTH-based and L-IDM-based models (and in most of the Gipps-based models).

Fig. 7 allows us also to compare model variants in terms of robustness. IDM-based models are the most robust, outperforming all other models regarding collision frequency (which is almost zero). Concerning all other models, the higher the model complexity the higher the collision frequency. Gipps-based models have been the second best. As to the linear controllers, the Gipps-equilibrium spacing policy has yielded more robust results than other spacing policies.

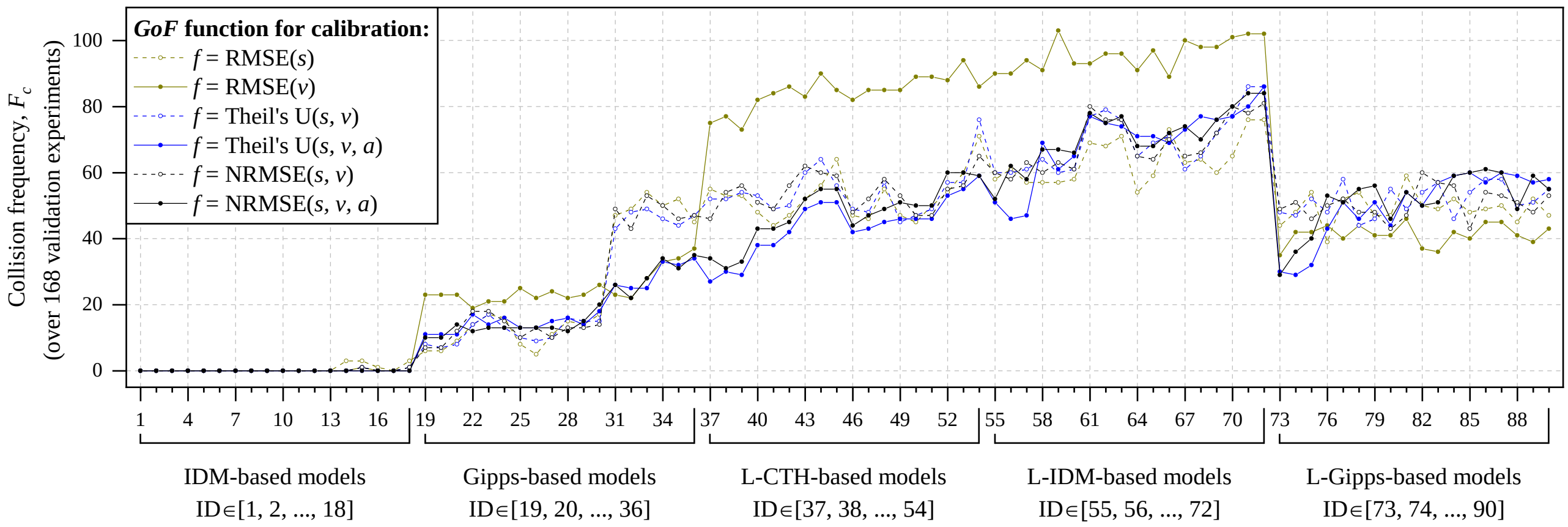

**Fig. 7. Collision frequency, for each model variant, over 168 validation experiments.**

## 6.2. Comparison of model variants

Table 5 reports the calibration (left half) and validation (right half) errors of the 90 models computed on spacing, speed and acceleration. Errors are expressed as percentage variation, from the base model, of the Root Mean Square Error. They measure the model ability to reproduce the spacing, speed, and acceleration dynamics observed in the data, relative to the base model performances. In each half of the table, also the *NRMSE*(*s*,*v*,*a*) percentage variation from the base model is reported. This measure represents a global validation score, as it sums up the errors on spacing, speed and acceleration.

Table rows are grouped by base model class. The values in each row of a group are the median percentage variations of the errors among all the calibration (left half) and validation (right half) experiments. Concerning validation results, it is worth noting that only validation experiments which were collision-free for all 18 models in a class have been selected for statistical analysis. Moreover, the rightmost column of the table shows the number of collisions that occurred in model validation, for each model. The numbers correspond to the collision frequencies depicted by the *NRMSE*(*s*,*v*,*a*) curve in Fig. 7.

Results show that, in general, physical extensions increase model accuracy in validation less than in calibration.

**Table 5**. Median calibration and validation errors of model variants, relative to the base models.

| Model class | Main extensions | ID | Calibration | [RMSE($s$)-base]/base | [RMSE($v$)-base]/base | [RMSE($a$)-base]/base | [$GoF$[a] -base]/base | Validation | [RMSE($s$)-base]/base | [RMSE($v$)-base]/base | [RMSE($a$)-base]/base | [$GoF$[a] -base]/base | # of collisions in validation |
|---|---|---|---|---|---|---|---|---|---|---|---|---|---|
| IDM-based models | Base | 1 | 28 calibration cases, % | - | - | - | - | 168 collision-free validation experiments[b], % | - | - | - | - | 0/168 |
| | | 2 | | 0.01 | 0.00 | 0.00 | 0.00 | | 0.01 | 0.00 | 0.00 | 0.00 | 0/168 |
| | | 3 | | 0.03 | 0.00 | 0.00 | 0.00 | | -0.01 | -0.01 | -0.01 | -0.01 | 0/168 |
| | LVD | 4 | | -1.19 | -6.85 | -7.10 | -6.33 | | -0.06 | -1.65 | -2.04 | -0.88 | 0/168 |
| | | 5 | | -1.19 | -6.80 | -7.11 | -6.38 | | -0.15 | -1.61 | -2.06 | -0.86 | 0/168 |
| | | 6 | | -1.17 | -6.42 | -6.02 | -4.89 | | -0.02 | -1.64 | -1.76 | -0.74 | 0/168 |
| | NVD | 7 | | -2.35 | -7.05 | -7.12 | -5.04 | | -1.29 | -5.07 | -4.20 | -4.34 | 0/168 |
| | | 8 | | -2.41 | -7.05 | -7.11 | -5.03 | | -1.14 | -4.99 | -4.21 | -4.24 | 0/168 |
| | | 9 | | -2.05 | -6.35 | -7.03 | -4.74 | | -1.99 | -5.44 | -4.41 | -4.82 | 0/168 |
| | PD | 10 | | -0.89 | -4.68 | -4.41 | -3.57 | | 0.24 | -0.74 | -0.55 | -0.25 | 0/168 |
| | | 11 | | -0.59 | -4.46 | -4.41 | -3.56 | | -0.02 | -0.92 | -0.58 | -0.32 | 0/168 |
| | | 12 | | -0.54 | -4.02 | -4.42 | -3.54 | | 0.13 | -1.03 | -0.79 | -0.30 | 0/168 |
| | PD+LVD | 13 | | -1.49 | -6.79 | -6.84 | -5.92 | | 0.19 | -1.83 | -1.60 | -0.43 | 0/168 |
| | | 14 | | -1.48 | -6.84 | -6.89 | -5.96 | | 0.31 | -1.63 | -1.42 | -0.22 | 0/168 |
| | | 15 | | -1.46 | -6.42 | -5.75 | -4.59 | | 0.33 | -1.54 | -1.56 | -0.25 | 0/168 |
| | PD+NVD | 16 | | -2.35 | -6.62 | -7.16 | -5.05 | | -0.74 | -4.77 | -3.44 | -3.35 | 0/168 |
| | | 17 | | -2.23 | -6.63 | -7.14 | -5.05 | | -1.19 | -4.89 | -3.48 | -3.34 | 0/168 |
| | | 18 | | -1.46 | -6.12 | -7.03 | -4.70 | | -1.00 | -5.09 | -3.80 | -3.74 | 0/168 |
| Gipps-based models | Base | 19 | 28 calibration cases, % | - | - | - | - | 121 collision-free validation experiments[b], % | - | - | - | - | 10/168 |
| | | 20 | | 0.05 | 0.02 | -0.03 | -0.01 | | 0.03 | 0.02 | -0.01 | 0.05 | 14/168 |
| | | 21 | | 0.09 | 0.01 | 0.00 | 0.00 | | 0.21 | -0.11 | -0.04 | 0.02 | 15/168 |
| | LVD | 22 | | -3.77 | -5.27 | -5.48 | -5.83 | | -0.96 | -3.32 | -3.53 | -3.89 | 14/168 |
| | | 23 | | -3.84 | -5.80 | -4.87 | -5.72 | | -1.69 | -3.88 | -3.91 | -4.13 | 14/168 |
| | | 24 | | -3.55 | -5.60 | -5.45 | -5.61 | | -1.50 | -3.04 | -3.88 | -3.93 | 15/168 |
| | NVD | 25 | | 0.47 | -4.13 | -3.63 | -3.02 | | -1.06 | -1.53 | -1.46 | -1.28 | 12/168 |
| | | 26 | | 0.48 | -4.89 | -3.73 | -3.09 | | -0.54 | -1.07 | -1.41 | -1.10 | 14/168 |
| | | 27 | | 0.92 | -4.22 | -3.58 | -3.05 | | -1.25 | -0.84 | -1.43 | -1.31 | 13/168 |
| | PD | 28 | | -7.71 | -9.24 | -7.15 | -10.63 | | -6.06 | -7.36 | -4.31 | -6.72 | 15/168 |
| | | 29 | | -8.39 | -9.30 | -7.09 | -10.60 | | -6.25 | -7.59 | -4.47 | -7.20 | 13/168 |
| | | 30 | | -7.51 | -9.19 | -7.13 | -10.62 | | -6.34 | -7.69 | -3.81 | -7.05 | 16/168 |
| | PD+LVD | 31 | | -7.96 | -9.34 | -4.68 | -8.97 | | -4.89 | -7.10 | -3.91 | -6.42 | 29/168 |
| | | 32 | | -9.90 | -9.28 | -5.70 | -9.00 | | -5.80 | -6.96 | -3.46 | -7.03 | 26/168 |
| | | 33 | | -8.16 | -8.79 | -5.95 | -9.01 | | -5.94 | -7.67 | -2.82 | -6.94 | 27/168 |
| | PD+NVD | 34 | | -12.91 | -8.51 | -5.57 | -8.00 | | -6.79 | -5.64 | -1.86 | -4.49 | 37/168 |
| | | 35 | | -12.02 | -8.14 | -5.31 | -8.25 | | -6.94 | -4.95 | -1.51 | -4.17 | 32/168 |
| | | 36 | | -11.00 | -8.05 | -5.94 | -8.34 | | -6.53 | -6.22 | -1.44 | -4.65 | 38/168 |
| L-CTH-based models | Base | 37 | 28 calibration cases, % | - | - | - | - | 98 collision-free validation experiments[b], % | - | - | - | - | 34/168 |
| | | 38 | | 0.02 | 0.03 | -0.01 | 0.00 | | 0.03 | 0.01 | -0.03 | -0.04 | 33/168 |
| | | 39 | | -0.01 | 0.04 | 0.00 | 0.00 | | -0.08 | 0.06 | 0.02 | 0.08 | 33/168 |
| | LVD | 40 | | -0.97 | -0.99 | -2.12 | -1.43 | | -0.32 | -1.26 | -2.52 | -1.47 | 45/168 |
| | | 41 | | -0.87 | -0.99 | -2.06 | -1.64 | | -0.36 | -1.23 | -2.69 | -1.46 | 45/168 |
| | | 42 | | -0.98 | -1.12 | -2.02 | -1.63 | | 0.05 | -1.15 | -2.86 | -1.77 | 45/168 |
| | NVD | 43 | | 3.66 | 1.84 | -0.35 | 0.40 | | 4.20 | 2.56 | 0.20 | 1.28 | 55/168 |
| | | 44 | | 3.39 | 2.00 | -0.36 | 0.40 | | 3.07 | 0.47 | -0.12 | 0.44 | 55/168 |
| | | 45 | | 3.73 | 1.88 | -0.26 | 0.78 | | 2.48 | 1.16 | 0.38 | 0.87 | 56/168 |
| | PD | 46 | | -1.05 | -0.91 | -1.92 | -1.64 | | -0.18 | -0.93 | -2.46 | -1.55 | 49/168 |
| | | 47 | | -1.05 | -0.93 | -2.07 | -1.89 | | 0.14 | -1.08 | -2.92 | -1.76 | 50/168 |
| | | 48 | | -1.16 | -0.89 | -2.07 | -1.87 | | 0.35 | -0.56 | -2.52 | -1.38 | 49/168 |
| | PD+LVD | 49 | | -1.20 | -1.54 | -2.36 | -1.57 | | -0.04 | -2.07 | -2.85 | -1.44 | 49/168 |
| | | 50 | | -1.24 | -1.60 | -2.31 | -1.74 | | -0.10 | -1.97 | -3.53 | -1.89 | 51/168 |
| | | 51 | | -1.27 | -1.46 | -2.32 | -1.77 | | 0.05 | -1.39 | -3.28 | -1.96 | 52/168 |
| | PD+NVD | 52 | | 4.36 | 1.96 | -0.14 | 1.19 | | 6.15 | 3.01 | 1.04 | 1.36 | 57/168 |
| | | 53 | | 4.63 | 2.36 | -0.41 | 0.90 | | 4.93 | 1.94 | -0.54 | 1.05 | 58/168 |
| | | 54 | | 4.54 | 2.17 | -0.33 | 1.08 | | 3.47 | 1.00 | -0.12 | 1.36 | 58/168 |
| L-IDM-based models | Base | 55 | 28 calibration cases, % | - | - | - | - | 76 collision-free validation experiments[b], % | - | - | - | - | 57/168 |
| | | 56 | | -0.07 | -0.17 | -0.07 | -0.02 | | -0.30 | -0.24 | -0.13 | -0.17 | 59/168 |
| | | 57 | | 0.01 | 0.04 | -0.06 | -0.01 | | -0.11 | -0.09 | -0.05 | -0.05 | 56/168 |
| | LVD | 58 | | -1.87 | -4.75 | -5.12 | -5.08 | | -0.68 | -3.56 | -4.29 | -2.62 | 65/168 |
| | | 59 | | -1.30 | -4.68 | -5.55 | -4.83 | | -0.70 | -4.45 | -5.33 | -3.90 | 67/168 |
| | | 60 | | -0.88 | -4.45 | -4.93 | -5.13 | | -0.25 | -3.50 | -4.87 | -2.86 | 68/168 |
| | NVD | 61 | | 2.53 | -1.50 | -3.00 | -1.06 | | 6.06 | 3.66 | 2.04 | 2.06 | 72/168 |
| | | 62 | | 1.14 | -0.72 | -2.78 | -0.85 | | 5.14 | 3.31 | 0.74 | 2.27 | 76/168 |
| | | 63 | | 1.63 | -0.70 | -3.14 | -0.29 | | 5.46 | 2.99 | 1.38 | 2.85 | 79/168 |
| | PD | 64 | | -1.14 | -3.06 | -3.65 | -2.80 | | -0.79 | -2.37 | -2.75 | -2.32 | 71/168 |
| | | 65 | | -1.68 | -2.74 | -3.24 | -2.96 | | -1.38 | -3.26 | -3.93 | -2.98 | 70/168 |
| | | 66 | | -1.63 | -3.68 | -3.19 | -3.38 | | -0.96 | -3.26 | -3.38 | -2.46 | 70/168 |
| | PD+LVD | 67 | | -1.37 | -5.20 | -6.35 | -5.50 | | -0.78 | -4.47 | -4.83 | -3.26 | 74/168 |
| | | 68 | | -2.15 | -5.66 | -6.22 | -5.51 | | -0.19 | -3.53 | -3.77 | -2.10 | 74/168 |
| | | 69 | | -1.94 | -5.52 | -6.22 | -5.32 | | 0.04 | -3.43 | -4.34 | -2.90 | 74/168 |
| | PD+NVD | 70 | | 0.98 | -0.62 | -3.07 | -0.71 | | 5.85 | 3.19 | 1.19 | 2.30 | 80/168 |
| | | 71 | | 1.35 | -1.61 | -3.39 | -0.61 | | 6.96 | 3.17 | 2.32 | 3.52 | 83/168 |
| | | 72 | | 0.99 | -0.89 | -2.72 | -1.27 | | 5.19 | 4.06 | 2.54 | 3.35 | 83/168 |
| L-Gipps-based models | Base | 73 | 28 calibration cases, % | - | - | - | - | 83 collision-free validation experiments[b], % | - | - | - | - | 35/168 |
| | | 74 | | -0.26 | 0.09 | 0.12 | -0.03 | | 0.02 | 0.03 | 0.27 | 0.11 | 32/168 |
| | | 75 | | -0.32 | -0.18 | 0.13 | -0.05 | | -0.10 | -0.24 | 0.05 | -0.10 | 38/168 |
| | LVD | 76 | | -1.26 | -3.42 | -3.83 | -3.70 | | 2.27 | -1.38 | -4.56 | -1.39 | 56/168 |
| | | 77 | | -0.74 | -3.03 | -3.72 | -3.74 | | -0.05 | -2.76 | -4.76 | -2.68 | 54/168 |
| | | 78 | | -0.36 | -3.26 | -3.28 | -3.49 | | 1.00 | -2.15 | -4.31 | -1.78 | 56/168 |
| | NVD | 79 | | 6.63 | 1.02 | -0.03 | 1.52 | | 6.35 | 0.87 | -0.03 | 1.77 | 54/168 |
| | | 80 | | 10.08 | 0.50 | -0.31 | 1.49 | | 2.43 | 1.10 | -0.59 | 0.45 | 45/168 |
| | | 81 | | 8.64 | 0.37 | 0.02 | 1.41 | | 4.76 | 2.26 | -0.19 | 1.22 | 50/168 |
| | PD | 82 | | -0.72 | -3.59 | -3.87 | -3.74 | | 1.14 | -1.96 | -3.53 | -1.66 | 56/168 |
| | | 83 | | -1.53 | -3.30 | -3.99 | -3.62 | | -0.17 | -2.84 | -4.13 | -2.02 | 55/168 |
| | | 84 | | -0.81 | -3.49 | -3.63 | -3.73 | | -0.15 | -2.62 | -4.69 | -2.19 | 60/168 |
| | PD+LVD | 85 | | -1.25 | -3.38 | -3.76 | -3.96 | | 0.62 | -2.76 | -4.41 | -2.17 | 62/168 |
| | | 86 | | -1.02 | -3.38 | -4.58 | -4.40 | | -0.31 | -3.88 | -5.17 | -2.80 | 61/168 |
| | | 87 | | -0.96 | -3.37 | -4.42 | -4.62 | | 0.31 | -4.72 | -5.35 | -4.00 | 57/168 |
| | PD+NVD | 88 | | 9.49 | 0.78 | -0.17 | 1.15 | | 3.22 | 1.80 | -0.61 | 1.61 | 53/168 |
| | | 89 | | 7.96 | 0.65 | -0.21 | 1.16 | | 5.01 | 0.46 | -1.46 | 1.00 | 59/168 |
| | | 90 | | 9.11 | 0.36 | -0.74 | 0.73 | | 6.27 | -0.46 | -1.03 | 1.69 | 60/168 |

[a] $GoF$ = NRMSE($s$,$v$,$a$). [b] The number of validation experiments, in which all 18 models from a model class do not lead to collisions in simulation.

This trend is exacerbated in the IDM-based models, and especially when linear dynamics, or perception delay, are added to the base model. The contributions of these extensions to the base model accuracy are less evident in validation than in calibration. In contrast to calibration results, nonlinear dynamics contribute to increasing model accuracy more than linear dynamics. Corresponding model variants retained the highest reduction of base model error.

Concerning the Gipps-based models, validation results confirm the trends, found in calibration, about the contribution of linear and nonlinear dynamics (the former increase base model accuracy more than the latter). Among variants, adding perception delay produces the highest increase of base model accuracy.

The linear controllers also show consistent results between calibration and validation. Linear dynamics were found to have a higher impact on accuracy than nonlinear ones. Concerning spacing policies, it is worth mentioning that the impact of physics-extensions on performances of L-CTH-based models between calibration and validation experiments was the same, i.e., the percentage variations of base model accuracy in validation were very close to the ones found in calibration.

# 7. Conclusion

A comparison framework has been developed to confront behavioural CF models and ACC algorithms, when augmented with detailed lower-level dynamics of increasing complexity, such as perception delays, vehicle dynamics, and vehicle/motor constraints. All possible combinations of five base models (CF and ACC) with different extensions, i.e., 90 different model variants, have been calibrated and validated against open-source trajectory data collected by the European Commission in 7 platooning experiments of four high-end vehicles controlled by ACC systems (Audi, Tesla, Mercedes, and BMW), following a platoon leader.

As to the models, the IDM and Gipps' behavioural CF models, and the ACC linear controller with a CTH spacing policy have been tested. Two new ACC algorithms have been also investigated, derived by coupling a linear controller with spacing policies from the IDM and Gipps' model. To the best of our knowledge, the proposed approach to enhance controllers through results from car-following theory is novel in the field.

The comparison of calibration results has allowed us to assess the contribution of each physical extension to the base model accuracy, that is the model ability to reproduce observed ACC vehicle trajectories. Secondly, we have compared performances of CF models and ACC algorithms, once augmented with the same extensions, and highlighted the differences emerged.

Eventually, since no true reason for preferring CF models or ACC algorithms to simulate ACC-equipped vehicles exists, we have quantified the accuracy of usual ACC algorithms to capture actual dynamics of commercial ACC vehicles. Ultimately, we have verified whether applying linear control laws – and not CF models – to simulate ACC vehicles, has significance in practice.

Vehicle models, calibrated against a trajectory of a vehicle measured in one experiment, have been also cross-validated against trajectories of the same vehicle from other experiments. Validation results have allowed us to quantify the transferability of calibrated parameter values to different car-following scenarios, and to measure the robustness of calibrated models in terms of vehicle collisions, a symptom of string instability. Moreover, by comparing model performances between calibration and validation experiments, we have outlined which physics-extension brings the highest contribution to model accuracy.

Results show that the IDM and Gipps CF models reproduce ACC vehicle dynamics more accurately than linear controllers with CTH spacing policy. Across all 90 model variants, *Gipps-based models* resulted the best performing ones, followed by the novel controller here proposed, i.e., the *linear controller* with *the Gipps' equilibrium speed-spacing function*, as spacing policy. The fact that Gipps' logics, originally conceived to interpret safe human drivers' behaviour, outperforms the approaches developed in the field to study ACC should be no surprise. In fact, commercial ACCs aim at mimicking human-like safe behaviors, and Gipps' is a safety-based model which converts an intuitive safe driving behavior in formulas.

Concerning transferability of calibrated model parameters, and model robustness to vehicle collisions, the IDM model, in all its variants, outperforms all other models, followed by Gipps-based models.

Results have also shown that using the $NRMSE(s,v,a)$ as objective function in CF/ACC model calibration sensibly improves model robustness and transferability of calibrated parameters. On one hand, this result is a confirmation of the recommendation provided in Punzo et al. (2021) about calibration settings and extends its validity to a considerable larger number of models (90 in this study) and to a larger dataset of ACC trajectories.

On the other hand, it proves that calibrating with the $NRMSE(s,v,a)$ enhance also model parameter robustness and transferability.

With respect to physics-extensions, calibration results have shown that adding linear vehicle dynamics to the base model largely improves model performances, and that linear dynamics are more effective than nonlinear ones. Adding perception delay also sensibly improves accuracy. In addition, the contribution of perception delay has the same order of magnitude of that produced by linear dynamics. This effect may be due to the fact that the actuation lag and the perception delay produce similar delayed vehicle dynamics, in simulation. However, the combined effect of adding both linear dynamics and perception delay is less than the sum of the single contributions.

Model validation results generally confirms the trends observed in calibration experiments. Exceptions are the IDM and Gipps' models. In validation, the IDM model has shown a considerable increase of accuracy with nonlinear dynamics in place of linear ones. Adding perception delay to Gipps' model, instead, has produced the highest increase of model accuracy, relative to the base model.

Overall, this study underlines that the research on adaptive cruise control can substantially benefit from traffic flow theory and can draw on it to develop more realistic models. This take is motivated by two facts, emerged from the work. From a modelling perspective, car-following models and ACC algorithms, as upper-level controllers, are proved to be conceptually equivalent, thus the replacement of actual logics is effortless. From a practical viewpoint, behavioral car-following models, and traffic flow theory in general, mimic the intelligence of human driving, which is the ultimate goal of automated driving. Therefore, traffic flow theory studies are crucial to accelerate both the development and the deployment of automated vehicle technologies. Surely, they can help avoiding unpleasant surprises, as that of string stability, a very well-known requirement of traffic which is currently neglected in commercial ACC systems (Ciuffo, 2021).

# Acknowledgment

The views expressed here are purely those of the authors and may not, under any circumstances, be regarded as an official position of the European Commission. The authors are grateful to Giovanni Albano, Aikaterini Anesaidou and Gaetano Zaccaria for the support provided during the work, as well as to Michail Makridis for the inspiring discussions. The authors would like to thank all the staff in AstaZero proving ground in Sweden for their support during the experiments. All the data used in the paper are part of the JRC openACC database openly available online. This research was partially funded by the Italian program PON AIM - Attraction and International Mobility, Linea 1 (AIM1849341-2, CUP E61G18000540007).